\documentclass[preprint2]{aastex61}
\shorttitle{$\varepsilon$ Indi B and C}
\shortauthors{Dieterich et al.}

\begin{document}

\title{DYNAMICAL MASSES OF $\varepsilon$ INDI B AND C: TWO MASSIVE BROWN DWARFS AT THE  EDGE OF THE STELLAR-SUBSTELLAR BOUNDARY}

\correspondingauthor{Sergio B. Dieterich}
\email{sdieterich@carnegiescience.edu}

\author{Sergio B. Dieterich}
\altaffiliation{NSF Astronomy and Astrophysics Postdoctoral Fellow}
\altaffiliation{Visiting Astronomer, Cerro Tololo Inter-American Observatory.
  CTIO is operated by AURA, Inc. under contract to the National Science Foundation.}
\affil{Department of Terrestrial Magnetism, Carnegie Institution \\
5241 Broad Branch Road, NW \\
Washington, DC 20015-1305, USA}

\author{Alycia J. Weinberger}
\affil{Department of Terrestrial Magnetism, Carnegie Institution \\
5241 Broad Branch Road, NW \\
Washington, DC 20015-1305, USA}

\author{Alan P. Boss}
\affil{Department of Terrestrial Magnetism, Carnegie Institution \\
5241 Broad Branch Road, NW \\
Washington, DC 20015-1305, USA}

\author{Todd J. Henry}
\altaffiliation{Visiting Astronomer, Cerro Tololo Inter-American Observatory.
  CTIO is operated by AURA, Inc. under contract to the National Science Foundation.}
\affil{RECONS Institute \\
 Chambersburg, PA, USA}

\author{Wei-Chun Jao}
\altaffiliation{Visiting Astronomer, Cerro Tololo Inter-American Observatory.
  CTIO is operated by AURA, Inc. under contract to the National Science Foundation.}
\affil{Georgia State University \\
  Atlanta, GA, USA}

\author{Jonathan Gagne}
\altaffiliation{NASA Sagan Fellow}
\affil{Department of Terrestrial Magnetism, Carnegie Institution \\
5241 Broad Branch Road, NW \\
Washington, DC 20015-1305, USA}

\author{Tri L. Astraatmadja}
\altaffiliation{Thompson Postdoctoral Fellow}
\affil{Department of Terrestrial Magnetism, Carnegie Institution \\
5241 Broad Branch Road, NW \\
Washington, DC 20015-1305, USA}

\author{Maggie A. Thompson}
\affil{Department of Astronomy \& Astrophysics, University of California, Santa Cruz \\
 Santa Cruz, CA, 95064, USA}

\author{Guillem Anglada-Escude}
\affil{School of Physics and Astronomy, Queen Mary University of London\\
London, UK}

\begin{abstract}
  We report individual dynamical masses for the
   brown dwarfs $\varepsilon$ Indi B and C, which have spectral types of
   T1.5 and T6, respectively, measured from astrometric orbit
  mapping. Our measurements are based on a joint analysis of
  astrometric data from the Carnegie Astrometric Planet Search and the
  Cerro Tololo Inter-American Observatory Parallax Investigation as
  well as archival high resolution imaging, and  use a Markov Chain Monte
  Carlo method. We find dynamical masses of 75.0$\pm$0.82 $M_{Jup}$
  for the T1.5 B component and 70.1$\pm$0.68 $M_{Jup}$ for the T6 C
  component. These masses are surprisingly high for substellar objects
  and challenge our understanding of substellar structure and
  evolution. We discuss several evolutionary scenarios proposed in the
  literature and find that while none of them can provide conclusive
  explanations for the high substellar masses, evolutionary models
  incorporating lower
  atmospheric opacities come closer to approximating our results. 
  We discuss the details of our astrometric model, its algorithm
    implementation, and how we determine parameter values via Markov Chain
    Monte Carlo Bayesian inference.

\end{abstract}

\section{Introduction} \label{sec:intro}
 The $\varepsilon$ Indi system (GJ 845, LHS 67) is a nearby triple system
  and its B and C components are amongst the T dwarfs closest to our
  Solar System.  It is a hierarchical system
comprising a K5V primary widely separated from a brown dwarf binary of
spectral types T1.5 and T6 for which we find
  an 11.4 year orbit. Some previously known properties of the system
are listed in Table \ref{tab:epsind}. The brown dwarf component was
first announced by \citet{Scholzetal2003}, who established common
proper motion to $\varepsilon$ Indi A at a projected separation of
402\farcs3. The B component was soon thereafter resolved as a close
binary with a projected separation of $\sim$0\farcs7 in VLT/NACO
observations by \citet{McCaughreanetal2004} and independently by
\citet{Volketal2003} on Gemini South.  \citet{Kasperetal2009} and
\citet{Kingetal2010} independently assigned spectral types T1.5 and T6
for the B and C components. Both studies note slightly sub-solar
metallicity.  We adopt [Fe/H] = $-$0.13$\pm$0.02 based on a weighted
mean of literature values for the A component listed in the PASTEL
Catalogue \citep{Soubiranetal2016}.

Our understanding of substellar structure and evolution is still
incomplete in several aspects. There are still large discrepancies
among the predictions of evolutionary models and the properties of
both individual objects and observed populations
\citep[e.g.,][]{Dieterichetal2014}. The atmospheres of these cool
objects are difficult to model due to molecule and cloud formation,
leading to a poor understanding of overall opacities, which in turn
affect the rate of cooling. Different assumptions regarding
atmospheric opacities have led to different evolutionary models
arriving at significantly different evolutionary rates as well as
different fundamental properties for objects at the stellar-substellar
boundary (Section \ref{sec:discussion}).
It has also been suggested that small changes in metallicity and cloud parameters,
which in stellar theory are considered secondary properties,
may play disproportionately large
roles in the structure and evolution of substellar objects
\citep{Burrowsetal2011}, thus further complicating attempts at a
general characterization. Our theoretical understanding must now be
constrained by observations of brown dwarfs amenable to extensive
characterizations that yield precise dynamical masses as well as
spectrophotometric properties such as metallicities and spectral
energy distributions. Significant progress has been made recently with
the publication of a large collection of substellar dynamical masses
\citep[e.g.,][]{Bowleretal2018,DupuyandLiu2017,
  DupuyandKraus2013,DupuyandLiu2012,Konopackyetal2010}.
However, most binary substellar systems
amenable to dynamical mass determinations in reasonable time scales
have very small projected separations that make a thorough
spectrophotometric characterization of the individual components
difficult. Even when individual spectra can be obtained, our poor
understanding of how spectral features vary with metallicity in these
cool atmospheres hinders the comparison of different systems.

\begin{deluxetable}{lcccc}
\tabletypesize{\scriptsize}
\tablecaption{Previously Known Properties of the $\varepsilon$ Indi System} \label{tab:epsind}
\tablehead{ \colhead{Property}  &
            \colhead{ A }       &
            \colhead{ B }       &
            \colhead{ C }       &
            \colhead{References\tablenotemark{a}} }
\startdata
Spectral Type        &      K5V          &          T1.5       &        T6         &     A: 1; B, C: 2, 3   \\
Parallax (mas)       &   276.06$\pm$0.28     &   \nodata           &    \nodata        &     4                  \\    
Distance (pc)        &   3.622$\pm$0.004     &   \nodata           &    \nodata        &     4                  \\
$\mu_{\alpha}$ (mas yr$^{-1}$)       &   3960.93$\pm$0.24    &   \nodata           &    \nodata        &     4                  \\
$\mu_{\delta}$ (mas yr$^{-1}$)      &   -2539.23$\pm$0.17   &   \nodata           &    \nodata        &     4                  \\
Separation (c. 2004) &   402\farcs3      &      0\farcs7       &      0\farcs7     &     5                  \\
$\text{}$[Fe/H]       &  $-$0.13$\pm$0.02  &      \nodata        &     \nodata       &     6                  \\
$V$                    &   4.68            &      24.12$\pm$0.03 &    $\geq$26.6     &     A: 1; B, C: 3      \\
$R$                    &   \nodata         &      20.65$\pm$0.01 &    22.35$\pm$0.02 &        3               \\
$I$                    &   \nodata         &      17.15$\pm$0.02 &    18.91$\pm$0.02 &        3               \\
$z$                    &   \nodata         &      15.07$\pm$0.02 &    16.53$\pm$0.02 &        3               \\
$J$                    &  2.89$\pm$0.29    &      12.20$\pm$0.03 &    12.96$\pm$0.03 &      A: 7; B, C: 3     \\
$H$                    &  2.35$\pm$0.21    &      11.60$\pm$0.02 &    13.40$\pm$0.03 &      A: 7; B, C: 3     \\
$K$                    &  2.24$\pm$0.24    &      11.42$\pm$0.02 &    13.64$\pm$0.02 &      A: 7; B, C: 3     \\
\enddata
\tablenotetext{a}{(1) \citet{Evansetal1957}, (2) \citet{Kasperetal2009}, (3) \citet{Kingetal2010}, (4) \citet{vanLeeuwen2007},
  (5) \citet{Scholzetal2003}, (6) \citet[][and references therein]{Soubiranetal2016}, (7) \citet{Cutrietal2003}}
\end{deluxetable}

The $\varepsilon$ Indi system's proximity to Earth (3.62 pc), its hierarchical
nature with a well known primary component, and the relatively short
period of the BC component make this system an ideal benchmark for the
study of substellar structure and evolution. In this paper we
determine dynamical masses for $\varepsilon$ Indi B and C by solving
the system's complete astrometric motion. We do so by measuring the motion
of the unresolved B-C system's photocenter using data from two separate observing
programs and then using high resolution adaptive optics images to scale
the photocenter's orbit to the individual barycentric orbits of the B
and C components.  We discuss our observations in Section
\ref{sec:obs}, our astrometric model in Section \ref{sec:model}, and
results in Section \ref{sec:results}. We conclude with a discussion of
how the dynamical masses we report constrain substellar models, with
particular emphasis on the stellar-substellar boundary in Sections
\ref{sec:discussion} and \ref{sec:conclusions}. Appendix \ref{appendixa}
describes the MCMC algorithm and its implementation in detail and
provides instructions for downloading and using it. Appendix \ref{appendixb}
provides a detailed example of deriving individual dynamical masses
from the photocenter's orbit.

\section{Observations} \label{sec:obs}
The complete characterization of the motion of an unresolved
astrometric binary system requires observations in which the motion of
the system's photocenter is measured with respect to the background of
distant stars as well as at least one epoch of resolved imaging
\citep{McCarthyetal1991, vandeKamp1968}. The photocenter is
  defined as the centroid of the point spread function of the combined
  light from both components, and its location lies along the
  separation vector between the two components.  The precise location
  of the photocenter along the separation vector is determined by the
  observed flux ratio of the primary to the secondary component, which
  is in turn a function of the components' intrinsic spectral energy
  distribution and the photometric filter used to perform the
  observations. The observed astrometric motions are the motions of
  the photocenter about the system's barycenter. These motions are
  not necessarily equivalent to the motion of the physical
  components. For a given binary star system the mass ratio and the
  flux ratio will generally not be the same, thus causing the
  photocenter and the barycenter to lie along different points in the
  system's separation vector. The photocenter thus traces an orbit
  about the barycenter that has the same orientation but is smaller
  than the orbits traced by either component about each other or about
  the barycenter. We discuss the scaling of these orbits and how they
  relate to dynamical masses in Section \ref{subsec:masses} and
  Appendix \ref{appendixb}.

Once the trigonometric
parallax motion, the proper motion, and the orbital motion of the
photocenter are deconvolved (Section \ref{sec:model}), the flux ratio
and separation obtained from a resolved image allows the scaling of
the photocenter's orbit to the physical orbits of the primary and
secondary components around the system's barycenter, and thus the
determination of individual dynamical masses via Kepler's Third
Law. We combined astrometric observations from the Carnegie
Astrometric Planet Search \citep[CAPS;][]{Weinbergeretal2016,
  Anglada-Escudeetal2012, Bossetal2009} and the Cerro Tololo
Inter-American Observatory Parallax Investigation
\citep[CTIOPI;][]{Henryetal2006, Jaoetal2005} to map the photocenter's
orbit, parallax, and proper motion. We then used archival high
resolution images taken with the Naos-Conica (NACO) imager on the Very
Large Telescope UT4 (VLT/NACO) adaptive optics system to determine the
scale factor between the photocentric orbit and the physical orbit. We
now discuss these data sets individually.

\subsection{CTIOPI Observations} \label{subsec:ctiopi}
The Cerro Tololo Inter-American Observatory Parallax Investigation is
a large astrometric program that began in 1999 on the
CTIO/SMARTS\footnote{Small and Moderate Aperture Research Telescopes, www.astro.gsu.edu/\~thenry/SMARTS}
Consortium 0.9-m telescope. The details
of the observing procedures and data reduction are discussed in
\citet{Jaoetal2005}. $\varepsilon$ Indi BC was observed on 33 nights
between 2004 and 2016. Typically, five exposures of 300 seconds each
were taken during each epoch, always in the Kron$-$Cousins $I$ band.  Exposures were usually taken within
half an hour of meridian transit to minimize differential color
refraction.  A single image taken with good seeing and low hour angle
was selected as the ``trail frame'' and compared to the 2MASS catalog
\citep{Skrutskieetal2006} to determine plate rotation and the plate
scale. The pixel coordinates of ten reference stars were then used to
create coordinate transformations linking each individual exposure to
the reference frame established in the trail frame. Table
\ref{tab:astro} lists the displacement of $\varepsilon$ Indi BC's photocenter with
respect to the first epoch of observation.  The uncertainties
correspond to the standard deviation of the several individual
exposures taken during an epoch.

\startlongtable
\begin{deluxetable*}{cccccccccc}
\tabletypesize{\scriptsize}
\tablecaption{Astrometric Observations \label{tab:astro}}
\tablehead{ \colhead{Night}     &
            \colhead{Program}       &
            \colhead{Jul Date}       &
            \colhead{$\alpha$ Displacement\tablenotemark{a}\tablenotemark{b}}       &
            \colhead{$\sigma_{\alpha}$}   &
            \colhead{$\delta$ Displacement\tablenotemark{a}}       &
            \colhead{$\sigma_{\delta}$}   &
            \colhead{$p_{\alpha}$\tablenotemark{c}}   &
            \colhead{$p_{\delta}$\tablenotemark{c}}   \\
            \colhead{     } &
            \colhead{     } &
            \colhead{     } &
            \colhead{mas}  &
            \colhead{mas}  &
            \colhead{mas}  &
            \colhead{mas}    }
\startdata
2004-07-30  &   CTIOPI  &     2453216.77  &  -1.55\tablenotemark{d} &   3.40  &  -4.36\tablenotemark{d} & 3.81  &   0.348  &   -0.578   \\
2004-08-07  &   CTIOPI  &     2453224.74  &           49.13  &          6.08  &         -65.45        &   1.93  &   0.227  &   -0.638   \\
2004-09-26  &   CTIOPI  &     2453274.61  &          389.74  &          8.48  &         -434.68       &   7.60  &   -0.531 &   -0.726   \\
2005-07-26  &   CTIOPI  &     2453577.77  &         3944.09  &          3.98  &       -2451.49        &   5.92  &   0.408  &   -0.541   \\
2006-05-18  &   CTIOPI  &     2453873.91  &         7330.16  &          3.95  &       -4247.16        &   3.55  &   0.943  &   0.267    \\
2006-07-05  &   CTIOPI  &     2453921.81  &         7785.40  &          3.10  &       -4751.86        &   5.02  &   0.679  &   -0.322   \\
2007-07-26  &   CTIOPI  &     2454307.82  &         11909.36 & 6.00\tablenotemark{e}  &   -7472.15    & 6.00\tablenotemark{e} &     0.413 &    -0.534  \\
2007-08-08  &   CTIOPI  &     2454320.78  &         12002.87 & 6.00\tablenotemark{e}  &   -7595.21    & 6.00\tablenotemark{e} &    0.219  &   -0.637   \\
2007-10-26  &   CTIOPI  &     2454399.52  &         12605.38 & 6.00\tablenotemark{e}  &   -8123.54    & 6.00\tablenotemark{e} &    -0.836 &   -0.535   \\
2009-07-30  &   CTIOPI  &     2455042.68  &         20045.24 &          1.12  &       -12700.78       &   1.88  &   0.353  &   -0.569   \\
2010-07-31  &   CTIOPI  &     2455408.81  &         24017.54 &          0.72  &       -15296.33       &  10.54  &   0.342  &   -0.576   \\
2010-09-29  &   CTIOPI  &     2455468.58  &         24432.31 &          4.04  &       -15751.05       &   2.68  &   -0.561 &   -0.711   \\
2011-07-01  &   CTIOPI  &     2455743.83  &         27741.27 &          0.78  &       -17493.49       &   2.76  &   0.731  &   -0.268   \\
2011-09-23  &   CTIOPI  &     2455827.62  &         28303.49 &          6.13  &       -18174.35       &   12.46 &   -0.477 &   -0.733   \\
2011-10-07  &   CTIOPI  &     2455841.60  &         28403.17 &          0.96  &       -18261.94       &   2.82  &   -0.653 &   -0.674   \\
2012-07-05  &   CTIOPI  &     2456113.82  &         31702.14 &          5.46  &         -19990.99     &   6.34  &  0.681   &  -0.324    \\
2012-09-12  &   CTIOPI  &     2456182.65  &         32155.83 &           1.77 &          -20569.75    &    6.44 &  -0.330  &  -0.751    \\
2012-10-25  &   CTIOPI  &     2456225.53  &         32489.34 &          11.83 &          -20792.76    &    7.19 &  -0.827  &  -0.536    \\
2013-07-12  &   CTIOPI  &     2456485.82  &         35684.01 &           9.78 &          -22522.56    &    4.46 &   0.601  &   -0.401   \\
2013-08-30  &   CTIOPI  &     2456534.67  &         36009.34 & 6.00\tablenotemark{e}  &  -22928.99    &   6.00\tablenotemark{e} &  -0.127 &   -0.741  \\
2013-10-15  &   CTIOPI  &     2456580.55  &         36319.18 &           7.41   &        -23202.31    &    3.59  &   -0.742  &  -0.622        \\
2013-10-18  &   CTIOPI  &     2456583.54  &         36362.57 &           2.17   &        -23215.94    &    6.49  &   -0.770  &  -0.600        \\
2014-09-03  &   CTIOPI  &     2456903.66  &         39979.74 &           3.33   &        -25407.18    &    1.87  &   -0.188  &  -0.749        \\
2014-10-18  &   CTIOPI  &     2456948.55  &         40304.05 &           5.08   &        -25680.00    &    5.82  &   -0.769  &  -0.603        \\
2014-10-27  &   CTIOPI  &     2456957.52  &         40390.31 &           9.04   &        -25723.07    &    1.06  &   -0.840  &  -0.526        \\
2015-06-02  &   CTIOPI  &     2457175.93  &         43230.20 &           7.99   &        -27026.47    &    8.55  &    0.926  &   0.087        \\
2015-07-15  &   CTIOPI  &     2457218.80  &         43607.49 &           3.14   &        -27456.49    &    3.04  &    0.566  &   -0.429       \\
2015-07-24  &   CTIOPI  &     2457227.79  &         43662.77 &           6.78   &        -27537.33    &    2.76  &    0.445  &   -0.518       \\
2015-10-26  &   CTIOPI  &     2457321.53  &         44341.67 &           7.56   &        -28191.90    &    4.16  &   -0.834  &  -0.538        \\
2015-10-30  &   CTIOPI  &     2457325.51  &         44372.36 &           1.41   &        -28201.98    &    10.33 &   -0.860  &  -0.500        \\
2016-08-13  &   CTIOPI  &     2457613.74  &         47773.37 &           1.62   &        -30204.06    &    5.60  &    0.131  &   -0.674       \\
2016-09-24  &   CTIOPI  &     2457655.59  &         48060.36 &           2.15   &        -30508.83    &    0.85  &   -0.507  &  -0.733        \\
2016-10-02  &   CTIOPI  &     2457663.59  &         48106.84 &           2.44   &        -30545.51    &    2.25  &   -0.609  &  -0.701        \\
\hline 
2007-07-05  &    CAPS   &     2454286.84  &   3.32\tablenotemark{d}  &   2.58   &  -3.46\tablenotemark{d} & 3.74 &     0.685 &    -0.319         \\
2007-09-01  &    CAPS   &     2454344.70  &         407.45     &         2.32   &        -512.72    &      2.78  &   -0.155  &  -0.744           \\
2008-07-16  &    CAPS   &     2454663.80  &         4115.67    &         2.53   &        -2642.71   &      4.34  &    0.543  &   -0.446          \\
2008-09-15  &    CAPS   &     2454724.69  &         4530.64    &         2.25   &        -3156.49   &      5.15  &   -0.380  &  -0.749           \\
2009-06-03  &    CAPS   &     2454985.89  &         7780.21    &         3.64   &        -4770.24   &      2.34  &    0.922  &    0.068          \\
2009-09-03  &    CAPS   &     2455077.66  &         8486.47    &         6.73   &        -5643.70   &      8.48  &   -0.194  &  -0.748           \\
2009-11-05  &    CAPS   &     2455140.59  &         8981.89    &         5.32   &        -6000.55   &      5.33  &   -0.893  &  -0.429           \\
2010-06-25  &    CAPS   &     2455372.82  &         11985.39   &         2.09   &        -7558.94   &      3.45  &    0.782  &  -0.203           \\
2010-07-27  &    CAPS   &     2455404.74  &         12225.44   &         2.56   &        -7872.91   &      3.41  &    0.399  &  -0.546           \\
2010-11-13  &    CAPS   &     2455513.52  &         13024.62   &         3.79   &        -8548.99   &      7.68  &   -0.917  &  -0.343           \\
2011-08-06  &    CAPS   &     2455779.73  &         16201.05   &         2.22   &        -10374.04  &      6.22  &    0.253  &  -0.623           \\
2011-10-03  &    CAPS   &     2455837.57  &        16583.30    &       2.71     &       -10776.09   &    5.28    &   -0.610  &   -0.697          \\
2012-07-29  &    CAPS   &     2456137.75  &        20063.05    &       4.15     &       -12702.65   &    6.20    &    0.362  &    -0.567         \\
2012-09-25  &    CAPS   &     2456195.61  &        20443.35    &       2.83     &       -13123.84   &    2.81    &   -0.519  &   -0.726          \\
2013-07-14  &    CAPS   &     2456487.81  &        23873.57    &       2.49     &       -14952.33   &    3.80    &   0.571   &   -0.423          \\
2013-08-15  &    CAPS   &     2456519.70  &        24094.03    &       2.16     &       -15232.39   &    2.50    &   0.105   &   -0.683          \\
2014-07-13  &    CAPS   &     2456851.84  &        27795.79    &       1.56     &       -17333.41   &    2.96    &   0.587   &   -0.411          \\
2014-08-18  &    CAPS   &     2456887.72  &        28039.64    &       6.00     &       -17649.62   &    4.74    &   0.061   &   -0.697          \\
2015-06-06  &    CAPS   &     2457179.84  &        31434.19    &       4.38     &       -19366.94   &    7.87    &   0.912   &   0.037           \\
2015-06-10  &    CAPS   &     2457183.88  &        31465.34    &       3.80     &       -19408.54   &    4.97    &   0.892   &   -0.014          \\
2015-07-27  &    CAPS   &     2457230.76  &        31837.50    &       3.19     &       -19854.96   &    5.19    &   0.401   &   -0.544          \\
2016-06-21  &    CAPS   &     2457560.86  &        35533.00    &       4.18     &       -21930.66   &    4.34    &   0.812   &   -0.161          \\
2016-08-12  &    CAPS   &     2457612.68  &        35917.19    &       3.16     &       -22418.54   &    3.54    &   0.148   &   -0.667          \\
2016-10-07  &    CAPS   &     2457668.55  &        36301.61    &       2.28     &       -22796.51   &    2.73    &   -0.665  &   -0.671          \\
\enddata
\tablenotetext{a}{Displacement measured relative to the first epoch of observation for the observing program. North and East are positive.}
\tablenotetext{b}{Angular displacement in normal coordinates, not in units of RA. See \citet{1997ESASP1200.....E}.}
\tablenotetext{c}{$p_{\alpha}$ and $p_{\delta}$ are the parallax factors at the given epoch.}
\tablenotetext{d}{The non-zero displacement for the first epoch of observation in each program is due to the small difference between 
         the mean displacement for the first epoch using all frames from that night and the individual frame chosen as the original reference for measuring 
         displacement. That frame is an arbitrary first epoch frame in the CAPS program and the ``trail frame'' for CTIOPI.}
\tablenotetext{e}{An uncertainty of 6.00 mas was adopted in epochs for which only one CTIOPI exposure was available.}
\end{deluxetable*}

\subsection{CAPS Observations} \label{subsec:caps}
The Carnegie Astrometric Planet Search has observed nearby low mass
stars since 2007 using a custom built astrometric camera (CAPSCam)
mounted on the Carnegie du Pont 2.5-m telescope at Las Campanas
Observatory. Technical aspects of CAPSCam are discussed in detail in
\citet{Bossetal2009}. \citet{Anglada-Escudeetal2012} and
\citet{Weinbergeretal2016} discuss details of the astrometric data
reduction. $\varepsilon$ Indi BC was observed for 24 epochs
distributed between 2007 and 2014. Typically 40
exposures of 60 seconds each were taken per epoch while the target was
within 1 hour of meridian transit. The pixel coordinates of
all bright stars in each frame were computed using centroiding
algorithms and the positions were used to create coordinate
transformations for each image. This procedure was iterated until the
most stable set of 28 reference stars was established.
As discussed in Section \ref{subsec:masses}, CAPSCam does not
  use a physical filter in the traditional sense. However the convolution of
  the Dewar window transmission and the detector response function approximates
  the $z$ band.

\subsection{ VLT / NACO Observations} \label{subsec:vlt}
We downloaded publicly available high resolution adaptive optics data
taken with the
NACO\footnote{http://eso.org/sci/facilities/paranal/instruments/naco.html}
instrument mounted on the European Southern Observatory's Very Large
Telescope (VLT) UT4. The data span several observing programs from
2003 to 2013. While the peaks of the PSFs of both components are
clearly visible in most images, the low Strehl ratio produces a wide
halo effect around each component.  The overlapping halos from both
components can shift the PSF centroids and make the separation between
the components appear smaller than it is in reality even at
separations a few times greater than the PSF's FWHM. To avoid this
effect, we used data only from the 2004 and 2005 observing seasons when
separations were close to maximum and the minimum flux measured in a
vector connecting both components was comparable to the mean sky flux.
The final adopted separations are the weighted averages of observations
  taken in the $J$, $H$, and $K_S$ bands. Strehl ratios were highly variable
  depending on the band and on how well the adaptive optics correction worked
during an individual exposure, and were generally less than 0.1.
As a check, we also measured separations using a six
parameter synthetic PSF fit and obtained only negligible differences
from the centroiding results.
We discuss the individual observations in Section \ref{subsec:masses}.

\section{The Astrometric Model} \label{sec:model}
 The full motion of each component of a binary system with respect to
 the sidereal frame of reference is the superposition of the system's
 proper motion, the parallax reflex motion, and the orbital motion
 about the system's barycenter. The mathematical formulation is
 derived in detail in many classical works on the astrometry of binary
 stars \citep[e.g.,][]{Hilditch2001, Heintz1978, vandeKamp1967}.
The same formalism applies in the case of the photocenter's
motion about the barycenter, which is what was observed in this study.
We therefore apply the following model to the orbit of the photocenter about the
barycenter and discuss how to obtain the physical orbits of the components
and dynamical masses in Section \ref{subsec:masses} and Appendix \ref{appendixb}.
We replicate the relevant equations here for reference. For an
unresolved astrometric binary, the displacement of the system's
photocenter is expressed as
\begin{eqnarray}
  \Delta \alpha = \mu_{\alpha} (t - t_0) + \Pi p_{\alpha} + (BX + GY), \\
  \Delta \delta = \mu_{\delta} (t - t_0) + \Pi p_{\delta} + (AX + FY),
\end{eqnarray}
where $\mu$ is the proper motion for each direction of motion, $t_0$ is the
epoch of first observation, $\Pi$ is the trigonometric parallax, and $p$
is the parallax factor for each direction of motion.
The last two terms in parenthesis denote the orbital motion. The
Thiele-Innes constants A, B, F, and G are defined in terms of
orbital parameters as
\begin{eqnarray}
  A = a (\cos \Omega \cos \omega - \sin \Omega \sin \omega \cos i), \\
  B = a (\sin \Omega \cos \omega + \cos \Omega \sin \omega \cos i), \\
  F = a (-\cos \Omega \sin \omega - \sin \Omega \cos \omega \cos i), \\
  G = a (-\sin \Omega \sin \omega + \cos \Omega \cos \omega \cos i).
\end{eqnarray}
Where $a$ is the semi-major axis, $\Omega$ is the longitude of the ascending node, $\omega$
is the longitude of periastron, and $i$ is the orbit's inclination.
$X$ and $Y$ are the elliptical rectangular coordinates defined as
\begin{eqnarray}
  X = \cos E - e, \\
  Y = (1 - e^2)^{1/2} \sin E
\end{eqnarray}
where $e$ is the eccentricity and $E$ is the eccentric anomaly, which
is related to the epoch of observation through Kepler's equation
\begin{equation}
  E - e \sin E = \frac{2 \pi}{P} (t - T)
\end{equation}
where $P$ is the orbital period and $T$ is the epoch of periastron passage.
The solution in the case of a single set of observations done with uniform
methodology is then a ten parameter problem: two components of proper
motion, the trigonometric parallax, and the orbital elements $a$, $P$,
$e$, $T$, $\Omega$, $\omega$, and $i$. As described below, three additional
parameters are needed to combine two data sets.
We generated posterior samples for each parameter using an MCMC algorithm and
  infer the value and uncertainty of each parameter from these samples (Section \ref{sec:results}).
We describe the MCMC algorithm in detail
in Appendix \ref{appendixa} while discussing the physical aspects of the model
here.

The displacements in Table \ref{tab:astro} are measured with respect
to the first epoch of observation for either the CAPS or CTIOPI data
set, with the onset of observations happening earlier for CTIOPI on
2004 July 30. We take that time as the time origin for proper motion
displacement and assume that the displacement of the system's
barycenter is linear and due solely to proper motion. At any given
time, the position of the system's photocenter relative to the barycenter is
the sum of the displacements due to trigonometric parallax and orbital
motion. We therefore subtract the parallax and orbital displacement of
the first epoch of observation from equations 1 and 2 so as to shift
the model to the data set's reference frame. For any epoch of
observation the displacements in Table \ref{tab:astro} can then be
modeled as

{\scriptsize
\begin{eqnarray}
  \Delta \alpha = \mu_{\alpha} (t - t_0) + \Pi p_{\alpha} + (BX + GY) - [\Pi p_{\alpha} + (BX + GY)]_{1^{\text{st}}\text{ epoch}}, \\
  \Delta \delta = \mu_{\delta} (t - t_0) + \Pi p_{\delta} + (AX + FY) - [\Pi p_{\delta} + (AX + FY)]_{1^{\text{st}}\text{ epoch}}.
\end{eqnarray}
}

The location of the system's photocenter relative to the physical
location of the two components in a binary system is dependent on the
components' flux ratio in a given photometric band. Because the CTIOPI
data were observed in the $I_{KC}$ band and the CAPSCam band
approximates the $z$ band (Section \ref{subsec:masses}) each data set
yields a different value for the semi-major axis.  We therefore treat
the semi-major axes as separate free parameters in the
astrometric model,
thus adding an extra parameter to the astrometric model by varying $a$
in equations 3 through 6 depending on the source of the observation
for a given epoch.

\subsection{Establishing the Sidereal Reference Frame and Zero Point Astrometric Corrections} \label{subsec:zeropoint}
The background stars used to establish the frame of reference are at
finite distances and therefore they also have small measures of
parallax and proper motion. In the case of trigonometric parallax all
reference stars have reflex motion in the same direction as the
science star and that causes the relative parallax to be slightly
offset from the trigonometric parallax corresponding to the star's
true distance. The so called relative to absolute parallax correction
is done using photometric distance estimates to the reference stars,
and is discussed in detail in \citet{Weinbergeretal2016} and \citet{Jaoetal2005}.
Because CTIOPI and CAPSCam
use different reference stars, their offsets should be
different. However, when fit independently, both systems give
consistent (to within 1$\sigma$) estimates of the parallax. Furthermore,
the offsets, 0.79\,mas for CTIPI and 0.1\,mas for CAPSCam, are smaller
than the uncertainty on the final joint parallax, 0.81\,mas.
The joint parallax determined from the MCMC is the
result of the best fit to all the data taken together, with no special
accounting for a possible parallax offset of one system with respect
to the other. Therefore the joint parallax posterior is already
broadened by any actual offset.  The best fit joint parallax also
agrees, without a correction, to that determined by Hipparcos
(276.06$\pm$0.28 mas, Table \ref{tab:epsind}) to
within our uncertainty.  Therefore, we proceed with our calculations
using the best fit joint parallax as our estimate of the true parallax.

The correction for proper motion is more difficult to realize because
unlike in the case of parallax, we cannot assume a general form for the
proper motion of the reference stars. Any proper motion measurement is
relative to the combined proper motion of the stars in the field, and
because the CTIOPI and CAPS reductions use different sets of reference
stars they have different zero point proper motion corrections. This
correction has no effect on the resulting trigonometric parallax or
orbit solution because it is entirely absorbed by the much larger
proper motion of the science star with reference to the background
field of reference. We therefore treat the proper motions for the two
data sets as free parameters and allow them to fluctuate individually
while the other parameters are solved jointly. At final count, the
astrometric model expressed in equations 10 and 11 then becomes a 13
parameter problem.

\section{Results} \label{sec:results}
\begin{figure}[h!]
  \includegraphics[scale=0.5]{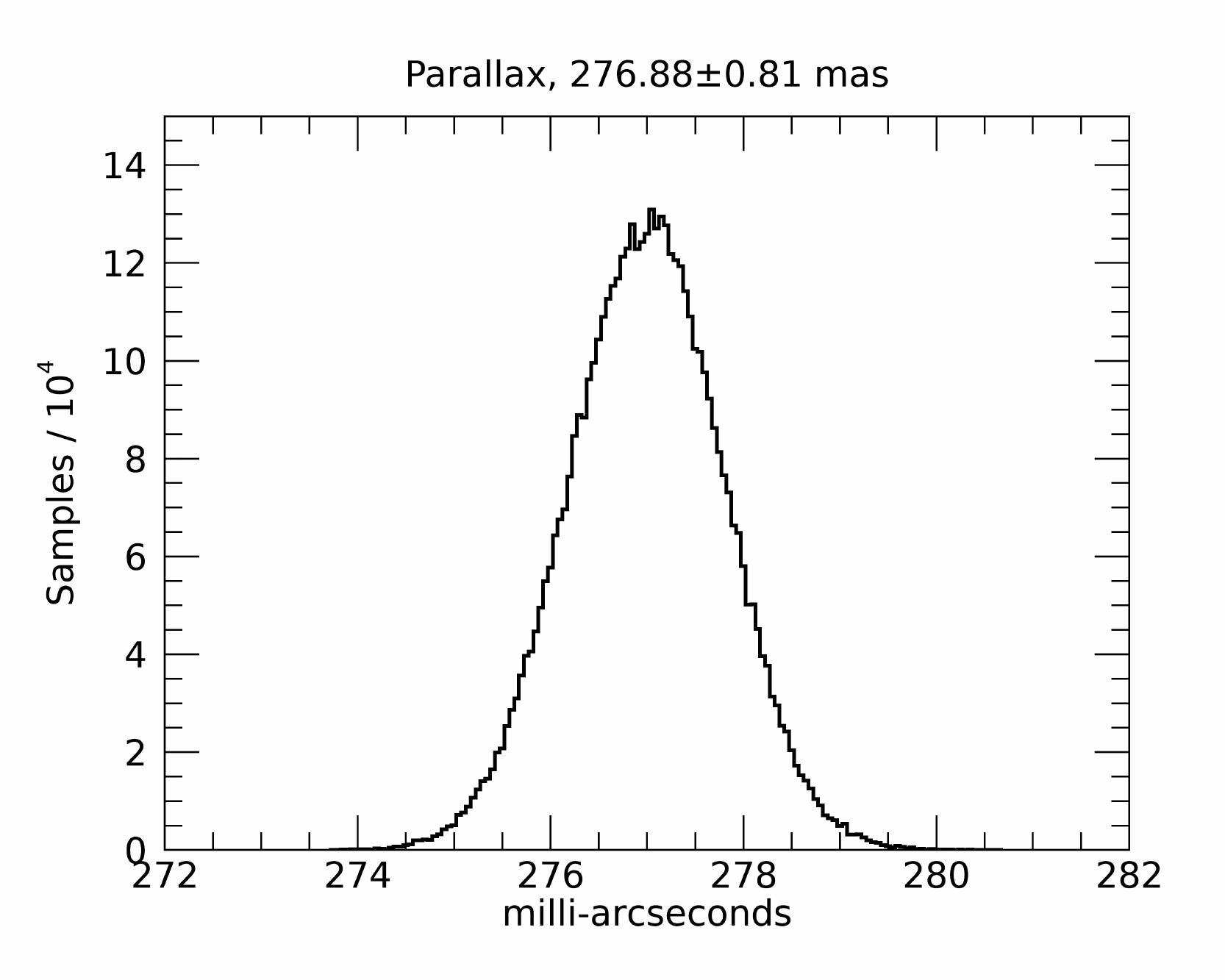}
  \caption{\scriptsize Histogram of posterior samples for trigonometric parallax derived using the
    combined CTIOPI and CAPS data. Quoted values are the median and standard deviation.
    The probability density function is based on
    5.2 million samples from 52 independent Markov Chains. \label{fig:pdf1}}
\end{figure}

\begin{figure*}[h!]
  \gridline{\fig{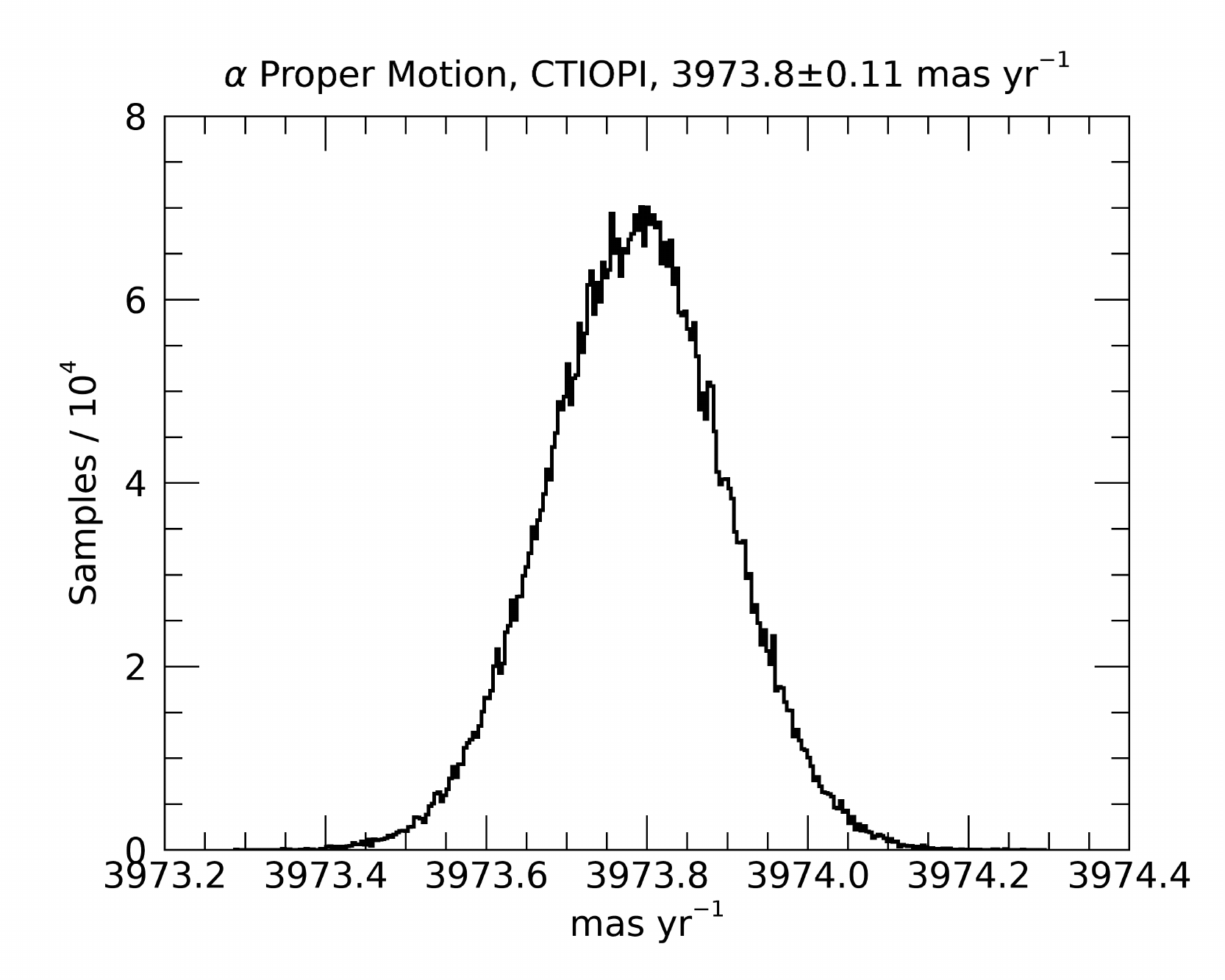}{0.4\textwidth}{}
            \fig{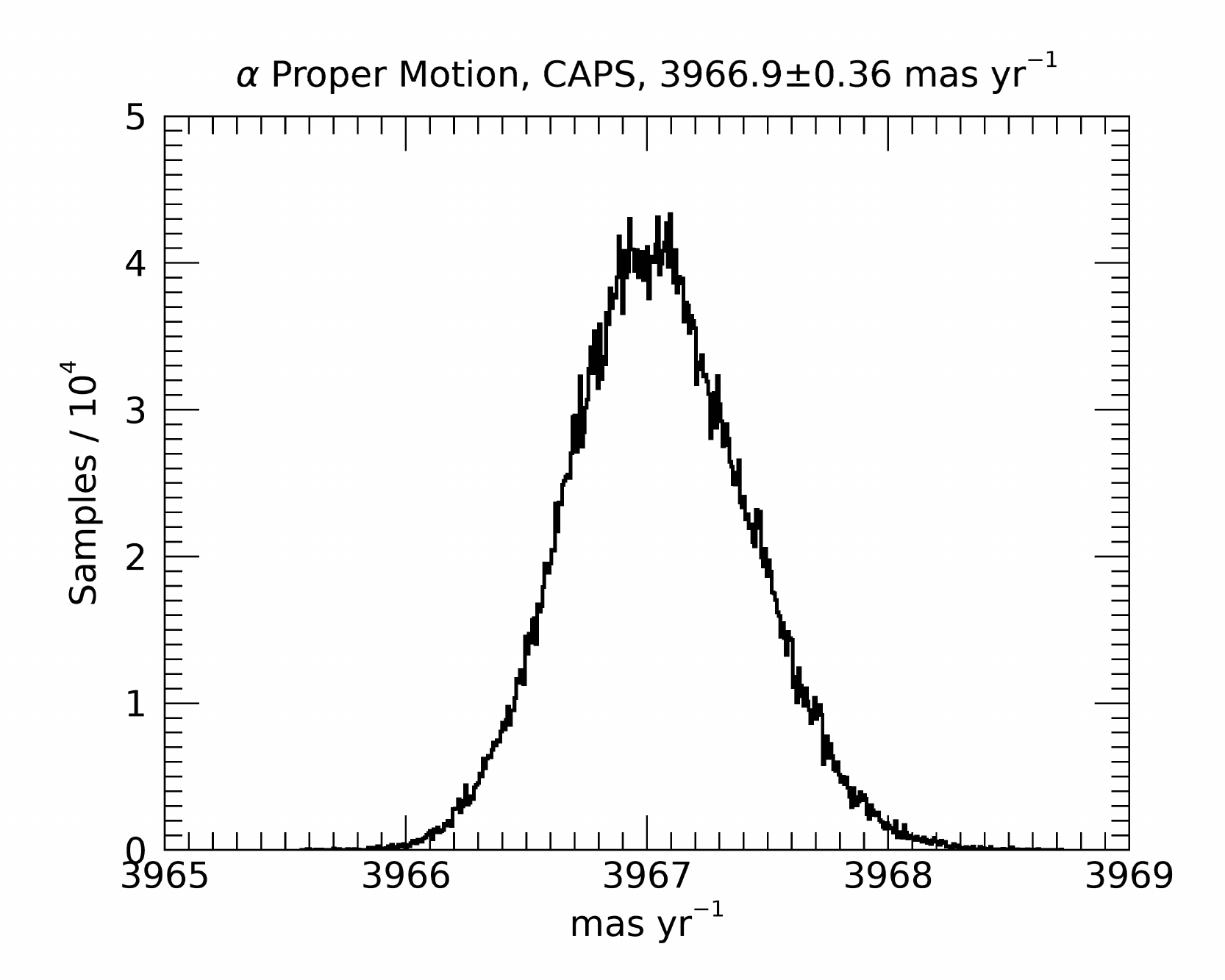}{0.4\textwidth}{}
           }
  \gridline{\fig{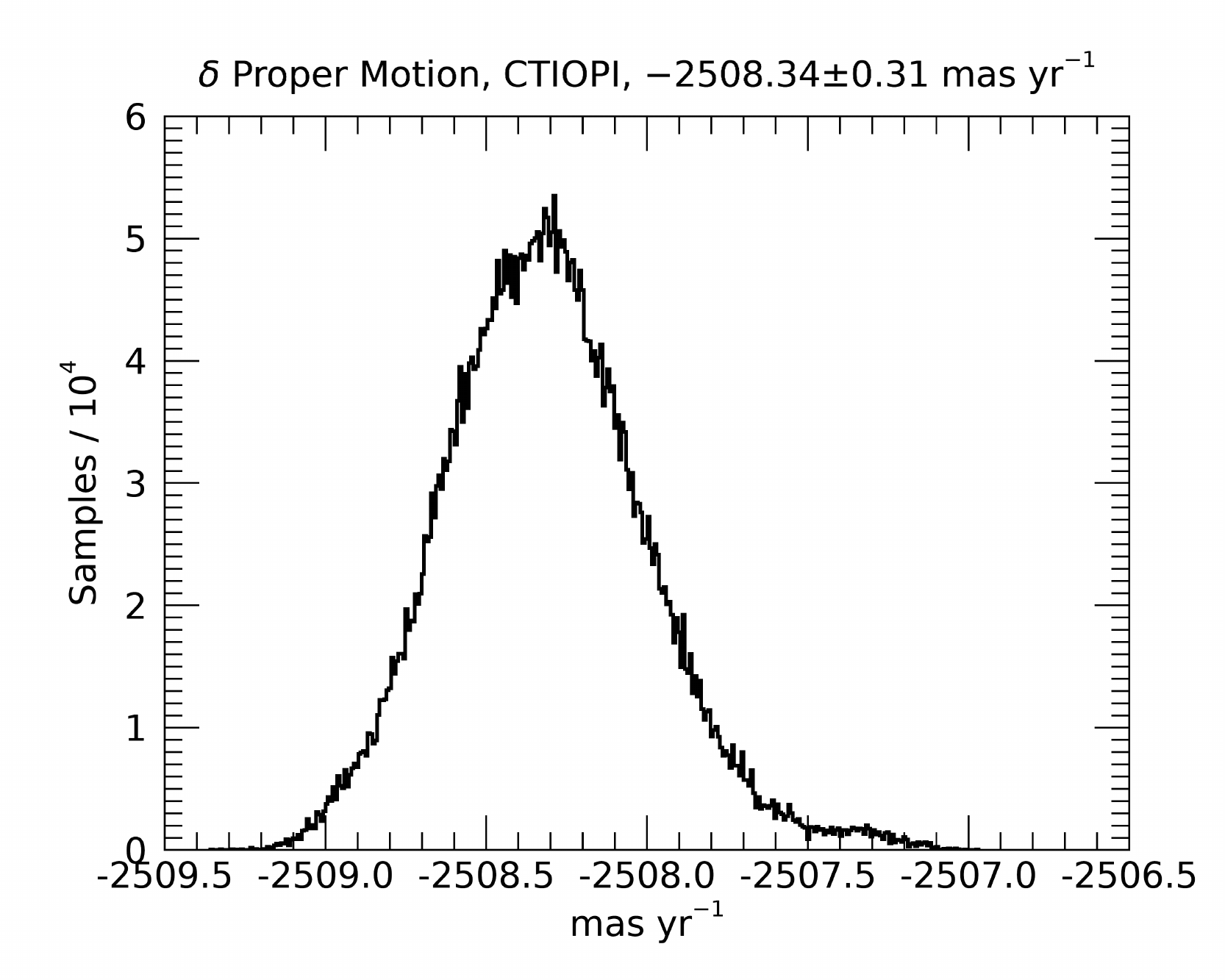}{0.4\textwidth}{}
            \fig{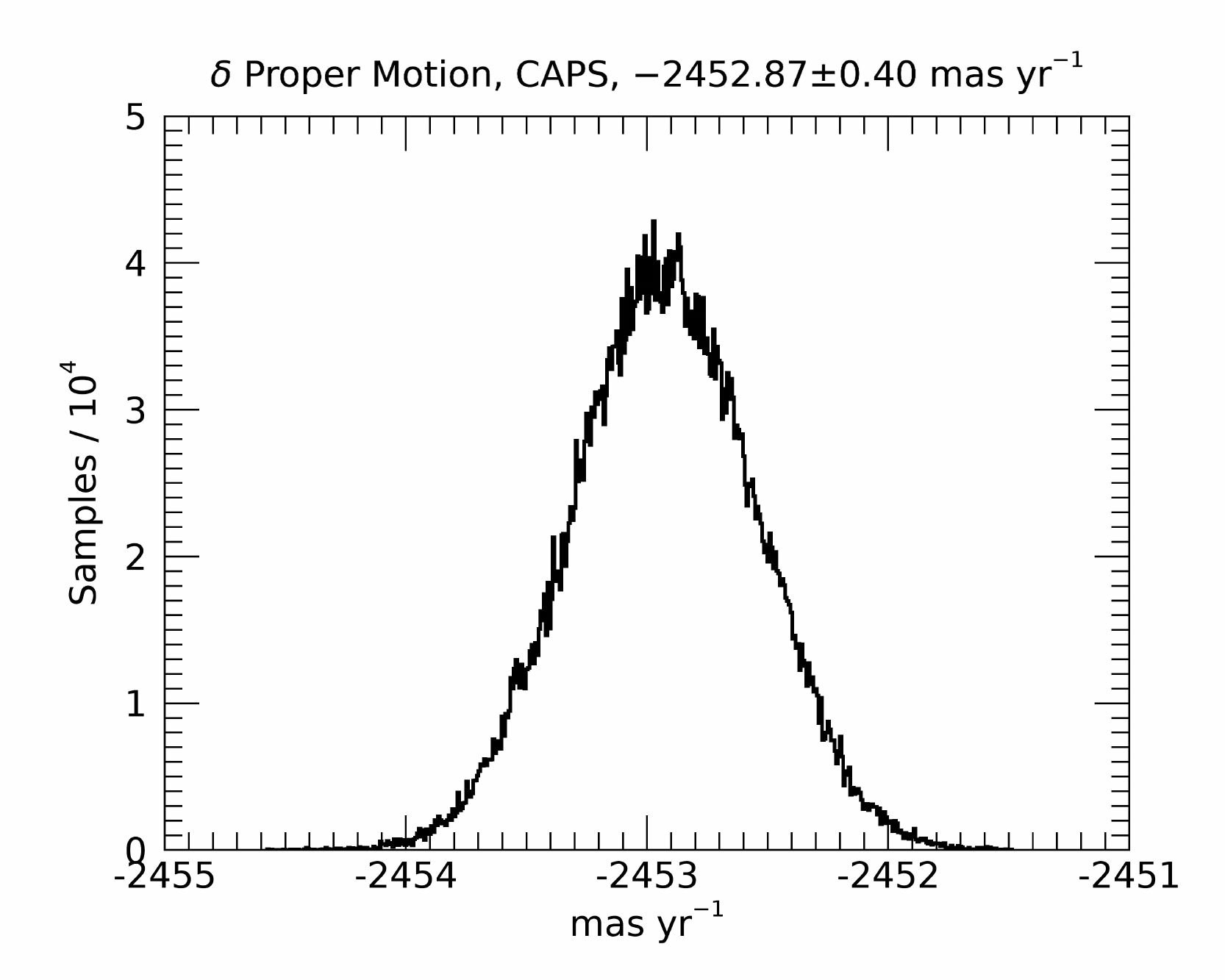}{0.4\textwidth}{}
           }
  \gridline{\fig{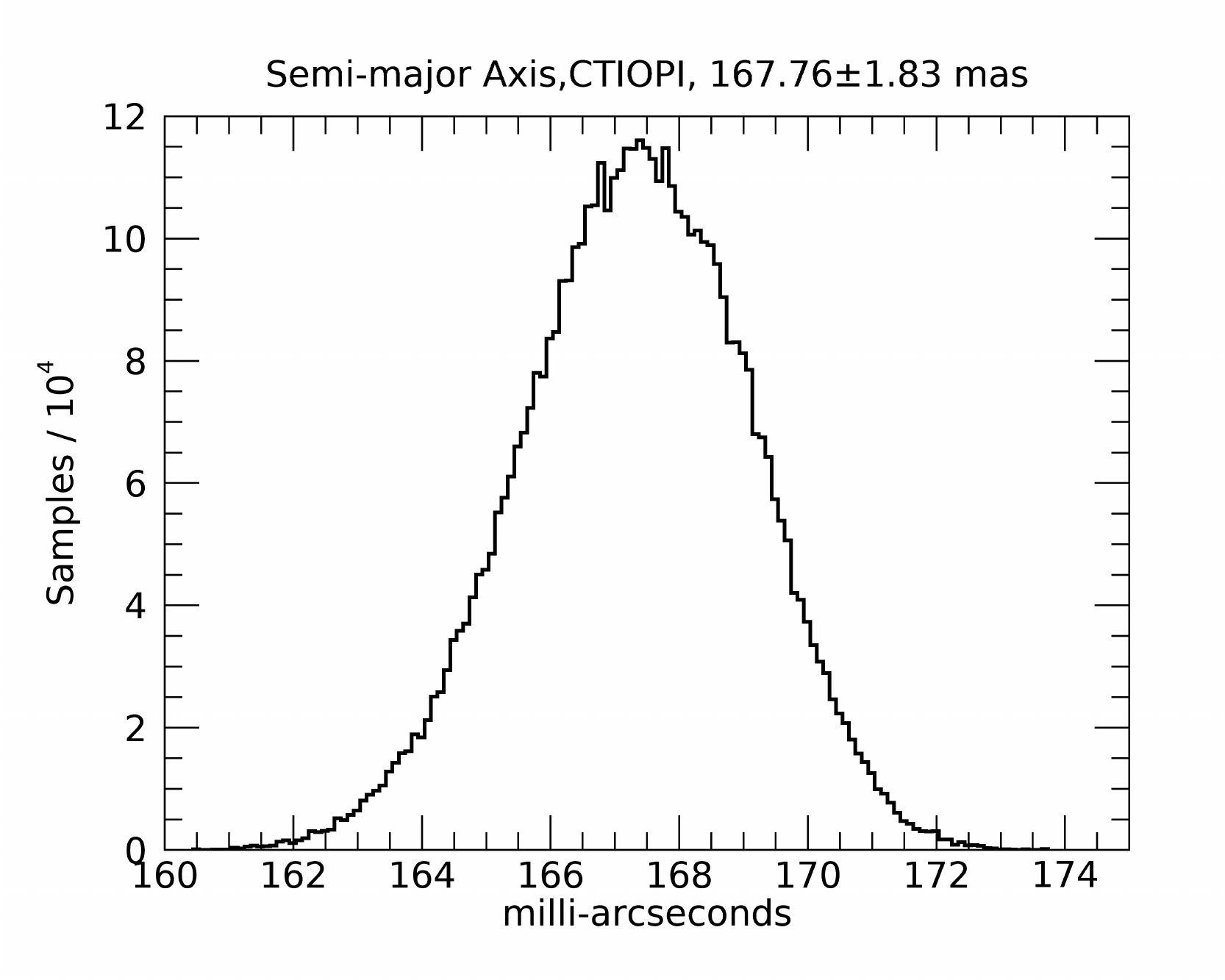}{0.4\textwidth}{}
            \fig{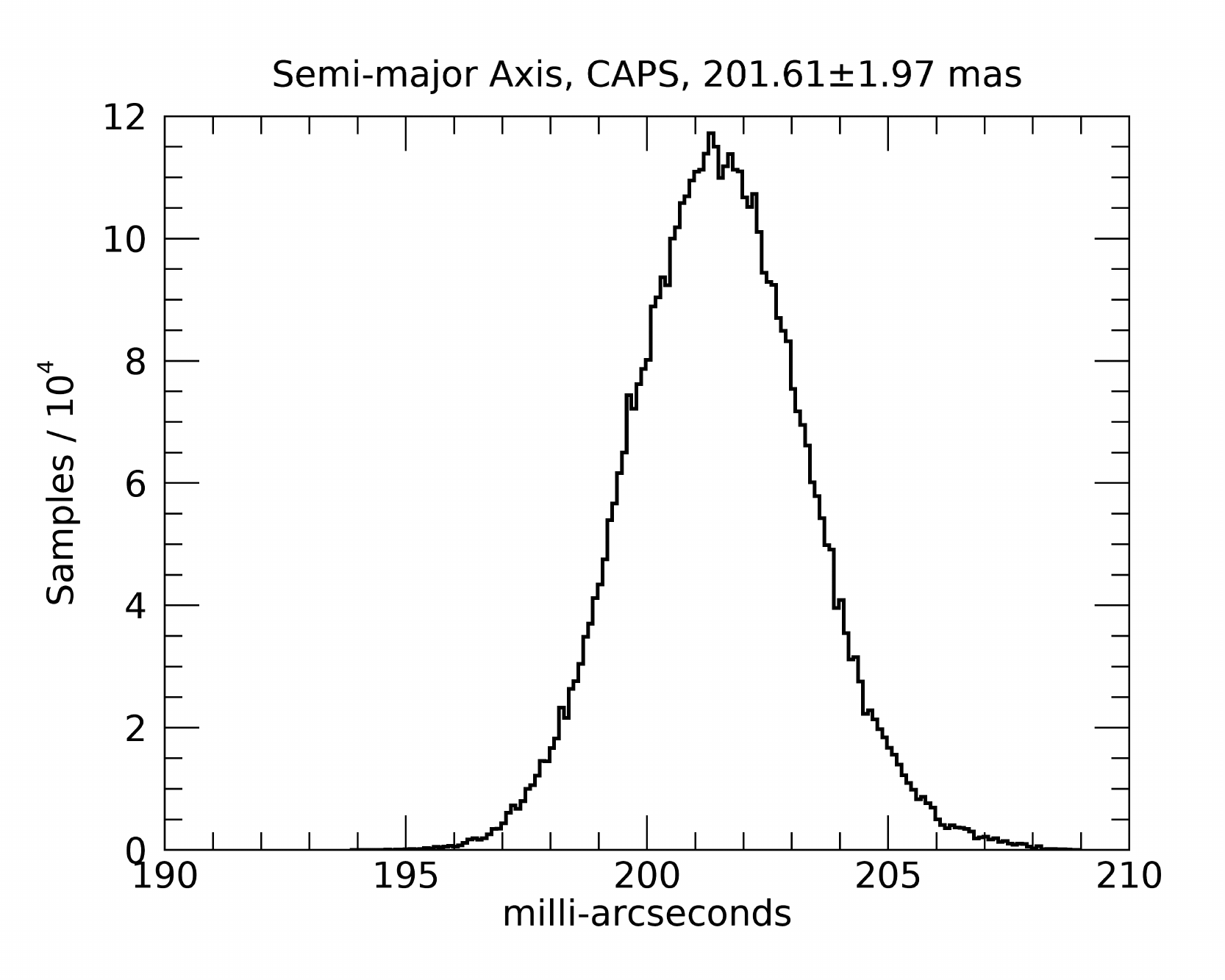}{0.4\textwidth}{}
           }
  \caption{\scriptsize Histograms of posterior samples estimating the probability density functions
    for parameters derived separately for the CTIOPI and CAPS data sets. Quoted values are the median
    and standard deviation for each parameter.
    The discrepancies in proper motion are addressed in Section \ref{subsec:zeropoint}.
    The photocentric semi-major axes are different due to the different filters used for CAPS and CTIOPI, as discussed
    in Section \ref{sec:model}. \label{fig:pdf2}}
\end{figure*}

\begin{figure*}[h!]
  \gridline{\fig{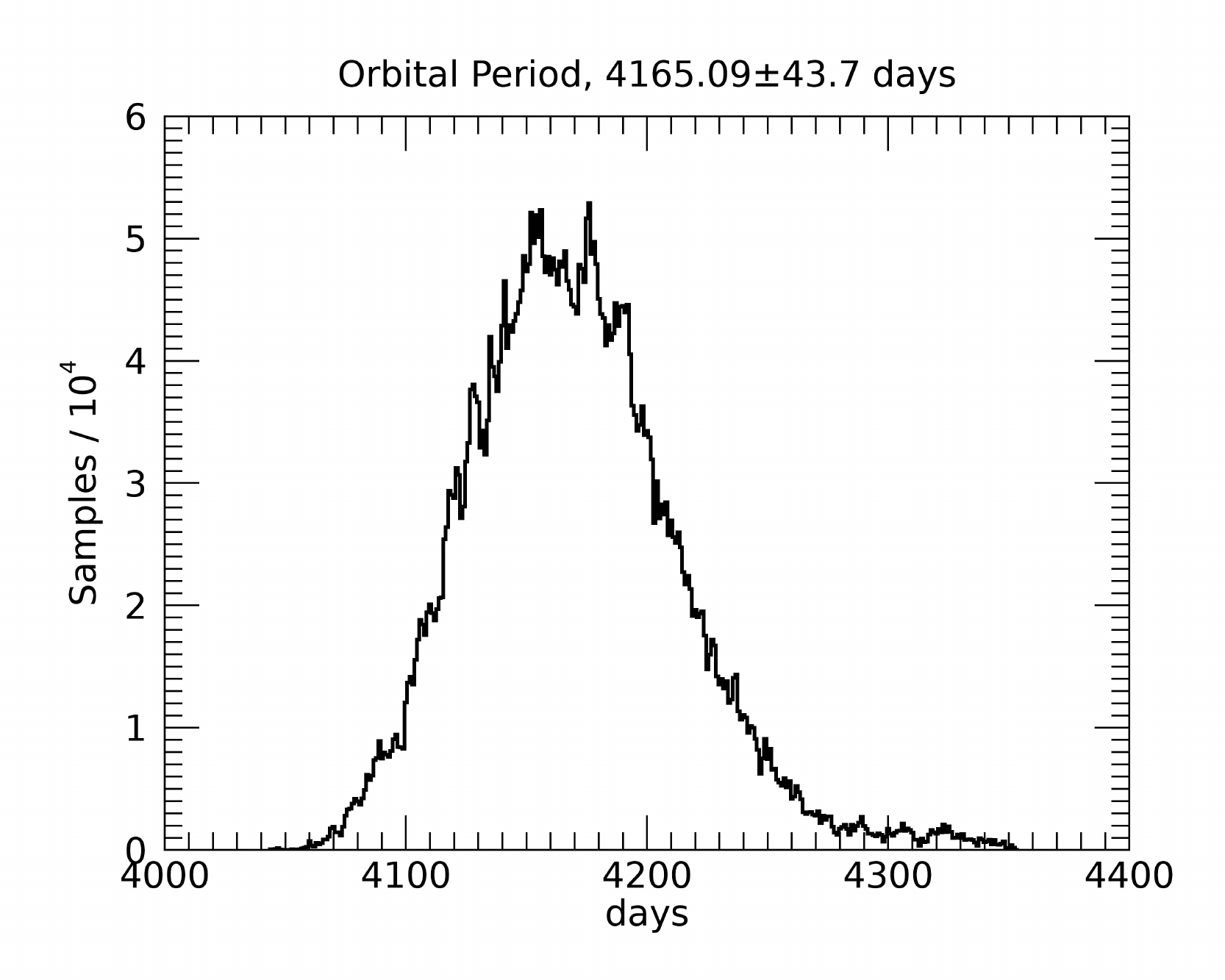}{0.45\textwidth}{}
            \fig{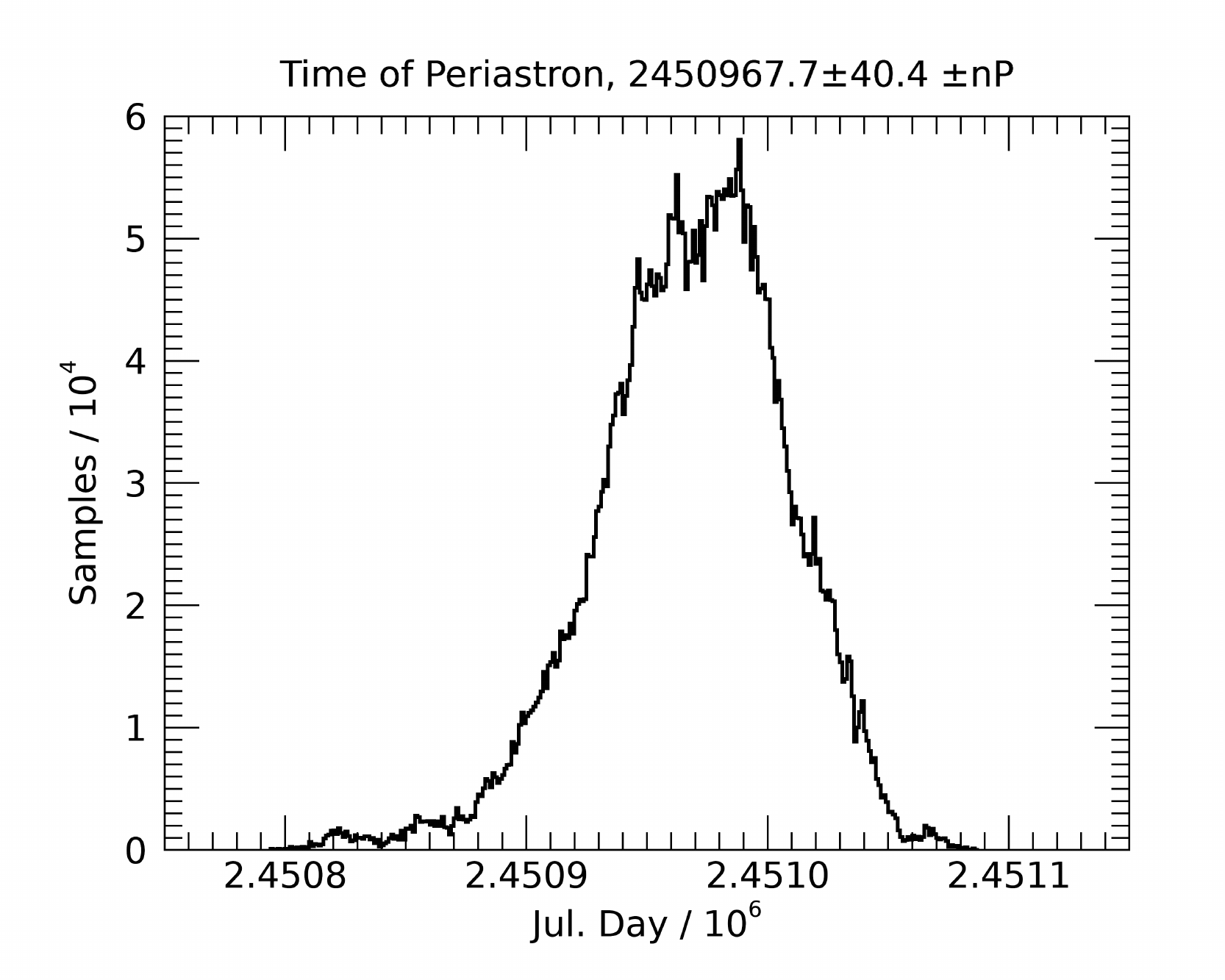}{0.45\textwidth}{}
           }
  \gridline{\fig{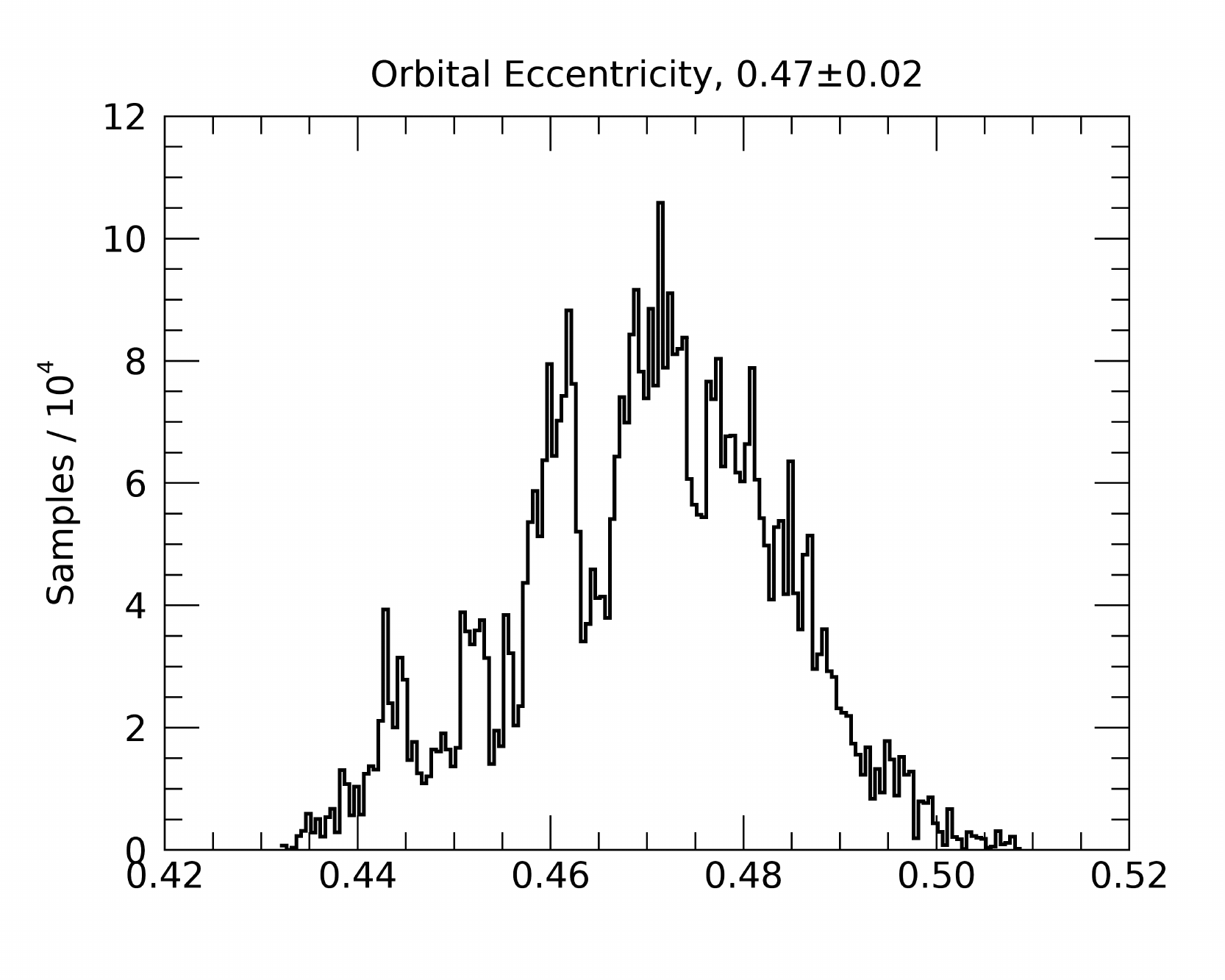}{0.45\textwidth}{}
            \fig{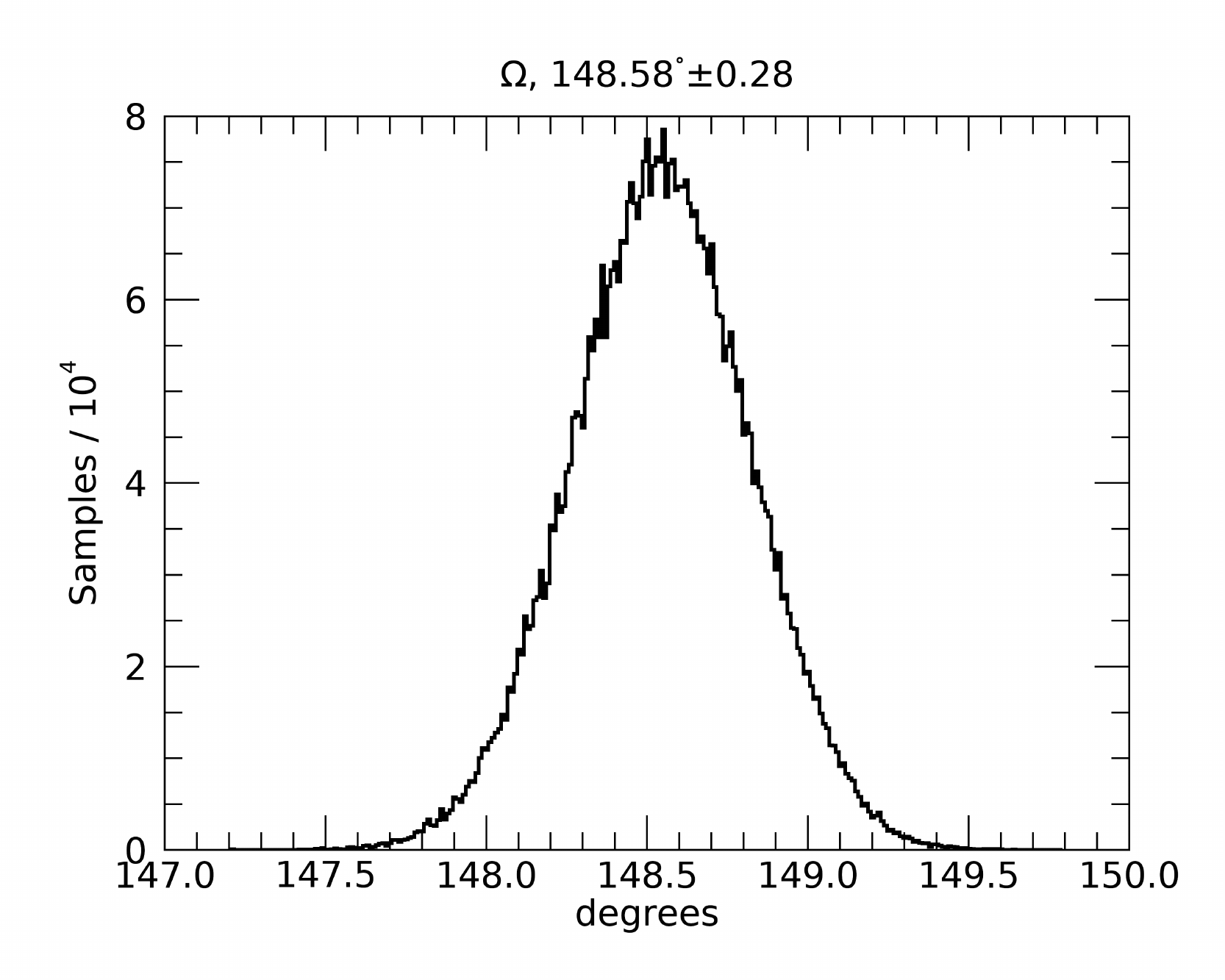}{0.45\textwidth}{}
           }
  \gridline{\fig{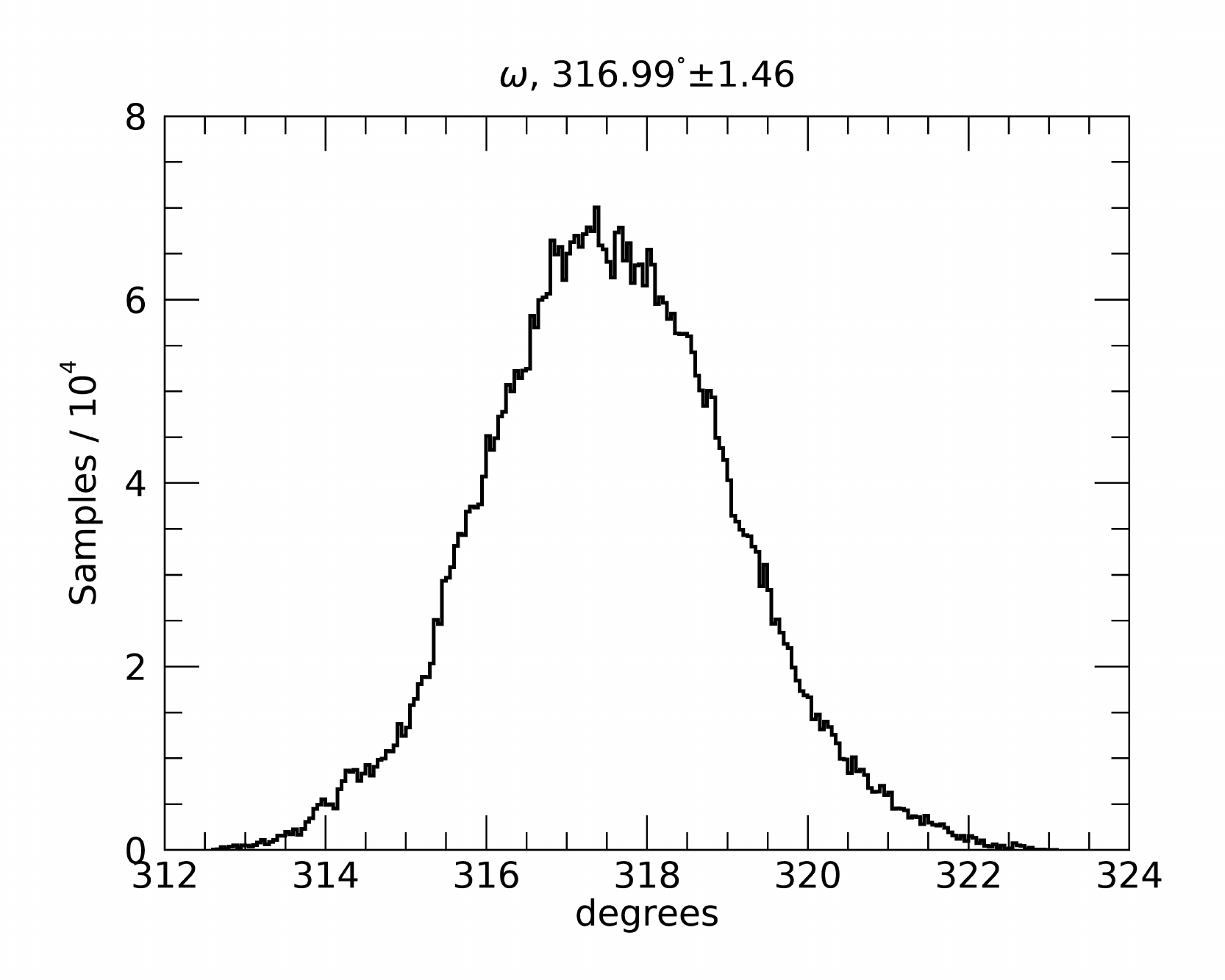}{0.45\textwidth}{}
            \fig{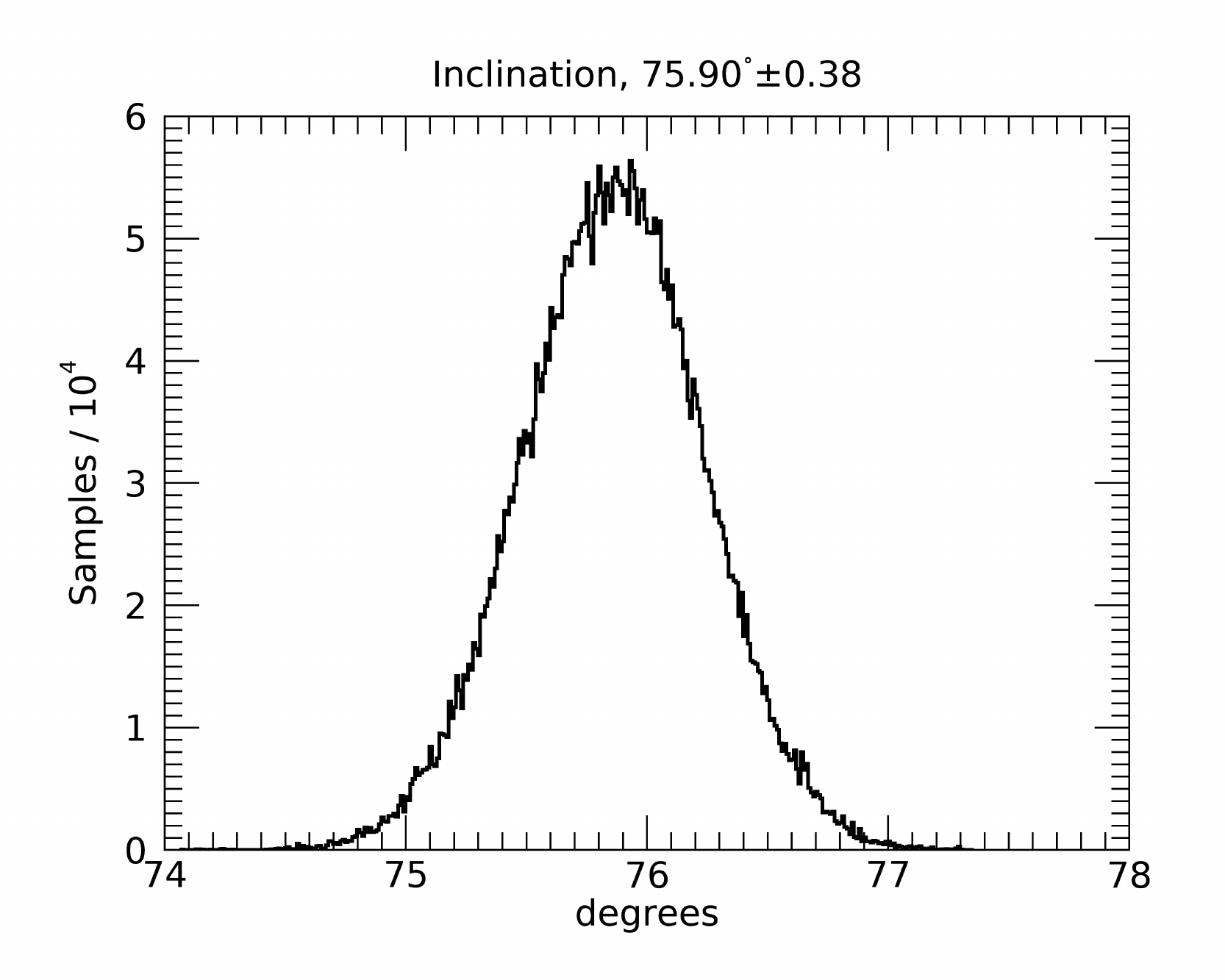}{0.45\textwidth}{}
           }
  \caption{\scriptsize Same as \ref{fig:pdf2}  for period, time of periastron passage, eccentricity, and orbit orientation angles
    derived using the combined CTIOPI and CAPS data.
    \label{fig:pdf3}}
\end{figure*}

Table \ref{tab:results} lists the astrometric parameters obtained from
the MCMC samples. The adopted values and their uncertainties
  are the medians and the standard deviations of the probability
  density functions, respectively. This approach is possible because
  all probability density functions are nearly Gaussian. Using the
  mean instead of the median values would produce differences that are
  negligible when compared to the uncertainties.  The probability
density functions for all parameters are shown in Figures
\ref{fig:pdf1}, \ref{fig:pdf2}, and \ref{fig:pdf3}. We describe
  some basic properties and numerical choices of our MCMC
  implementation here and provide a general description of the
  algorithm in Appendix \ref{appendixa}.  Section
  \ref{subsec:statistics} presents statistical tests and discusses the
  convergence of the Markov chains.

All probability density functions are based on the last 100000 steps
of 52 independent 2 million step chains, thus
comprising a total of 5.2 million MCMC samples
per parameter. No chain thinning was applied.  Uniform priors
covering parameter intervals much wider than what is physically
possible given the observations were used for all parameters
(Section \ref{spider}). The step scale was determined in an
  adaptive manner according to Equation A2. The process creates a
  distribution of step sizes centered about a chosen step value for a
  chosen scaling parameter. We chose the scaling parameter to be
  trigonometric parallax and the central step value to be 1 mas
  because that is the typical uncertainty in our astrometric
  measurement.  We then randomly divided or multiplied this central
  value (1 mas) by an uniformly generated random number between 0 and
  10 to create a broad distribution. The other parameters were scaled
  according to equation A3 so as to vary in a nearly random fashion on
  small scales ($\lesssim$ 100 steps) while causing the variations of
  all parameters to have effects of nearly the same magnitude in the
  large scale of the overall probability density function. This
  mechanism prevents parameters that heavily influence the overall
  astrometric motion, such as the very high proper motion, from also
  dominating the MCMC convergence at the expense of other parameters.

\begin{deluxetable}{rccc}[t!]
\tabletypesize{\scriptsize}
\tablecaption{Astrometric Results\tablenotemark{a} \label{tab:results}}
\tablehead{ \colhead{Parameter}     &
            \colhead{Value}         &
            \colhead{1$\sigma$ uncertainty}   &
            \colhead{Units}     }
\startdata 
$\Pi$                            &            276.88  &              0.81             &        mas                        \\ 
$d$                             &            3.61    &           $^{+0.016}_{-0.015}$    &        pc                         \\
Relative CTIOPI $\mu_{\alpha}$    &           3973.80  &             0.11              &        mas yr$^{-1}$              \\
Relative CTIOPI $\mu_{\delta}$    &           -2508.34 &             0.31              &        mas yr$^{-1}$              \\
Relative CAPS $\mu_{\alpha}$      &            3966.99 &              0.36             &        mas yr$^{-1}$              \\
Relative CAPS $\mu_{\delta}$      &           -2452.87 &             0.40              &        mas yr$^{-1}$              \\
$a_{CTIOPI}$                      &             167.76 &               1.83            &        mas                        \\
$a_{CAPS}$                        &             201.61 &               1.97            &        mas                        \\
$P$                               &            4165.09 &              43.7             &        day                        \\
$e$                               &               0.47 &                0.02           &        \nodata                    \\
$T$                               &          2450967.7$\pm n$P &      40.4              &        JD                         \\
$T$                               &            1998.45$\pm n$P &      0.11             &        epoch                      \\
$\Omega$                          &             148.58 &               0.28            &        degree\tablenotemark{b}    \\
$\omega$                          &             316.99 &               1.46            &        degree\tablenotemark{b}    \\
$i$                               &              75.90 &                0.38           &        degree    \\
\enddata
\tablenotetext{a}{All quantities refer to the system's photocenter.}
\tablenotetext{b}{Measured from North through East.}
\end{deluxetable}

Our joint solution yields
a trigonometric parallax of 276.88$\pm$0.81\,mas, which is in
excellent agreement with the Hipparcos parallax for the A component:
276.06$\pm$0.28\, mas. We therefore detect no depth separation between
the A and BC components.

\begin{figure*}[ht!]
  \includegraphics[scale=0.4]{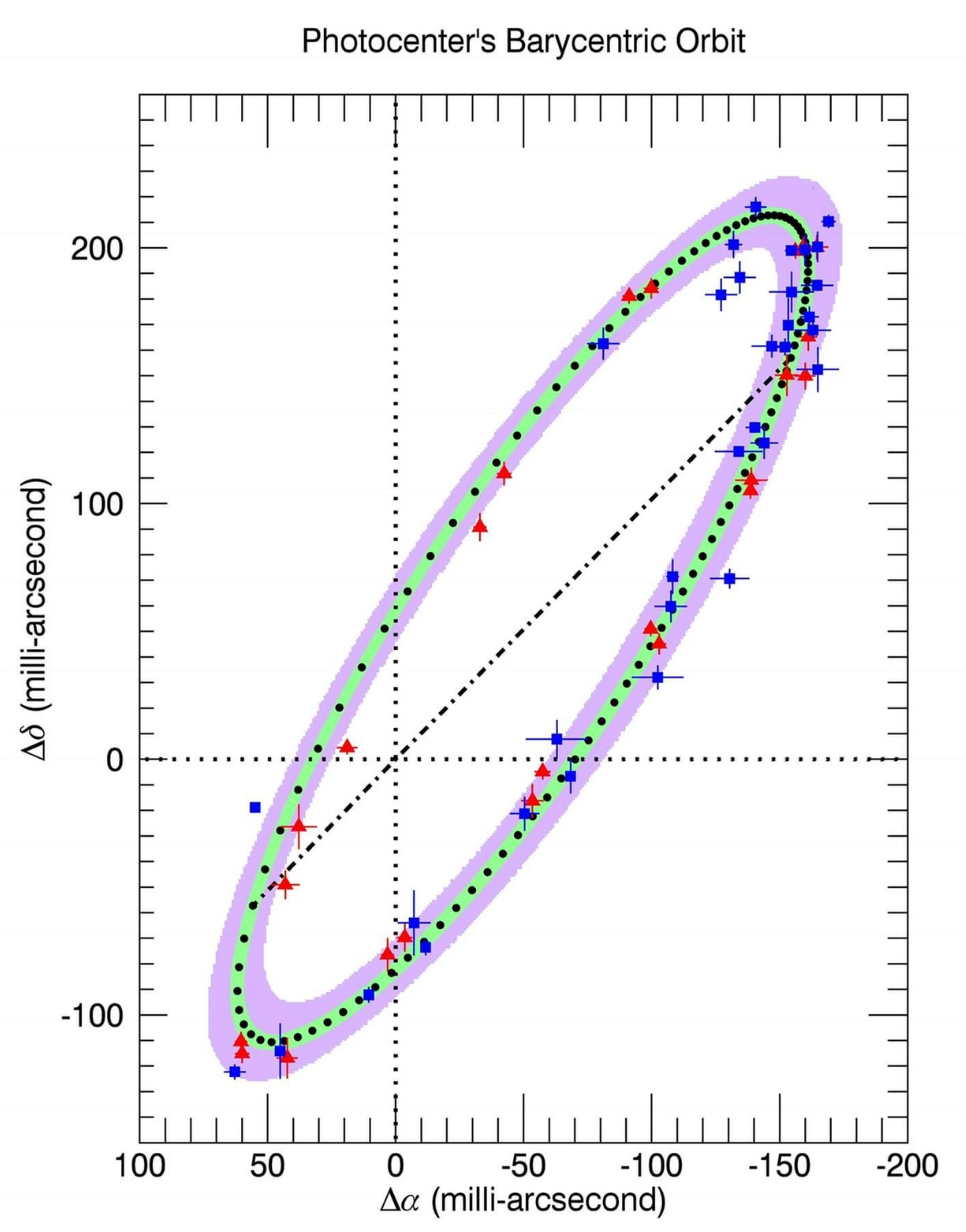}
  \caption{\scriptsize Sky projection of the barycentric orbit of
    $\varepsilon$ Indi BC's photocenter. The filled black circles
    denote time intervals of approximately 40 days. The shaded
    contours indicate the 1$\sigma$ (green) and 3$\sigma$ (purple)
    uncertainties on the projected orbit. The observed data (Table
    \ref{tab:astro}) are over-plotted with the proper motion and
    parallax subtracted. Blue squares indicate CTIOPI data and red
    triangles indicate CAPS data. The CTIOPI data was scaled up to match
    the semi-major axis of the CAPS data for clarity. The dash-dot line is the projection
    of the orbit's major axis.} \label{fig:orbitcontour}
\end{figure*}

Figure \ref{fig:orbitcontour} illustrates the orbit solution and the observed displacements
with the proper motion and parallax solutions subtracted. The shaded
contours indicate the 1$\sigma$ and 3$\sigma$ uncertainties of the
orbit solution based on a Monte Carlo simulation of 10000 possible
orbits given the values and uncertainties in Table
\ref{tab:results}. The semi-major axis obtained for the CTIOPI $I_{KC}$
band data (167.76$\pm$1.83 mas) was scaled up to match the semi-major
axis for the CAPS data (201.61$\pm$1.97 mas) for clarity. The 
smaller
semi-major axis in the $I_{KC}$ band reflects the trend towards bluer
colors for later T dwarfs
in the infrared (Table \ref{tab:epsind}), which decreases the flux
difference between the B and C components.  The formal solution shows
excellent agreement with the data, with 40 out 52, or 77 percent, of
the observations overlapping the 1$\sigma$ uncertainty contour,
indicating that the formal uncertainties may be slightly
overestimated.

The integrated light photometric variability of the BC component in
the CTIOPI data is 15.9\,mmag in the $I$ band. This result is the
standard deviation of the BC component's flux, measured by aperture
photometry, over all epochs when
compared to the sum of the flux of all reference stars, excluding any
found to be variable to more than 5\,mmag.
This value is
significantly smaller than the 136\,mmag inferred by \citet{Koen2013}.
Koen notes that his variability data appear to be correlated with
seeing and that such correlation is a clear indication of systemic
error. He concludes that the true variability is likely smaller than
his formal value.  The semi-major axis of the photocenter's orbit is a
function of both displacement and flux ratio and therefore the
uncertainty in the semi-major axis can serve as a check on
variability. For both data sets, our uncertainties in the semi-major
axis are approximately 1 percent (Table \ref{tab:results}), therefore
suggesting that the variability must be of that order or smaller.
We note, however, that the astrometric
observations provide only sporadic time coverage and do not rule
out isolated variations in flux as high as the ones noted in
\citet{Koen2013}. The lower variability is consistent with other
studies indicating that for field aged T dwarfs variability is generally in the
order of a few percent \citep[e.g.,][]{Radigan2014}. Variability data
are not available from the CAPSCam data set.

\begin{figure*}
\gridline{\fig{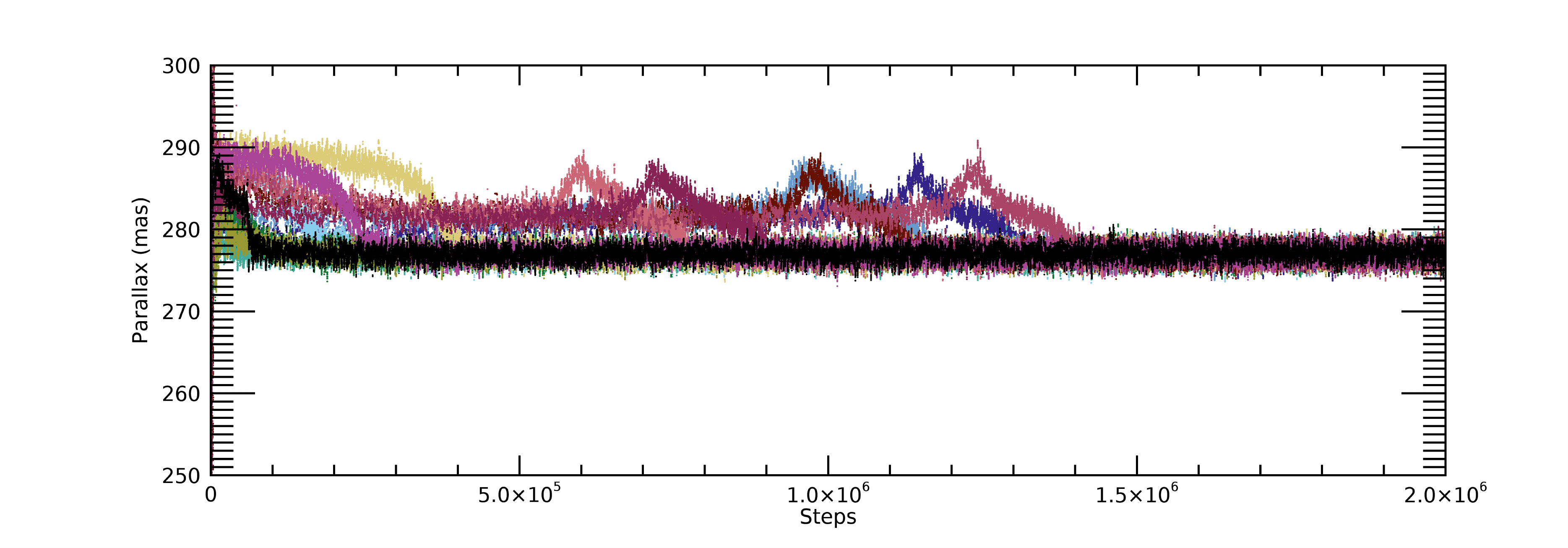}{0.9\textwidth}{}
          }
\gridline{\fig{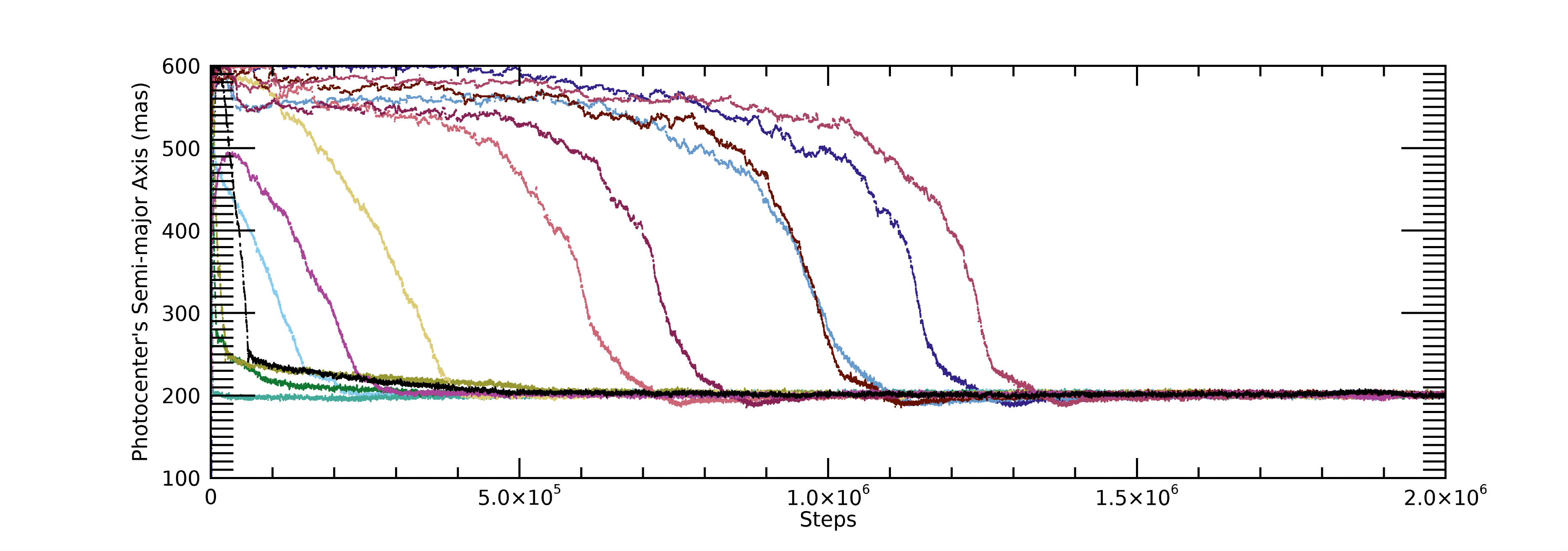}{0.9\textwidth}{}
          }
\gridline{\fig{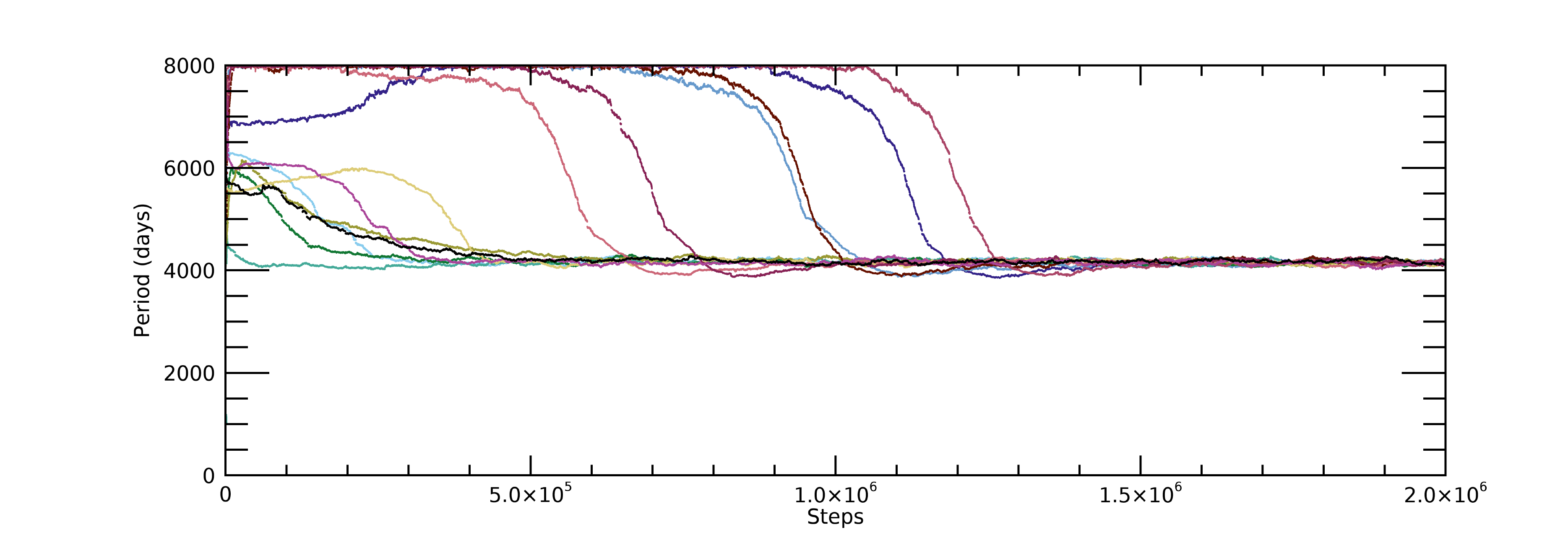}{0.9\textwidth}{}
          }
\caption{\scriptsize Full two million step Markov chains for the three
  parameters used in mass determination: trigonometric parallax,
  the semi-major axis of the photocenter's orbit in the CAPS data,
  and the orbital period.  Only 13 out of 52 chains are plotted
  for clarity. The vertical axes show the full range in which each
  parameter was allowed to fluctuate, essentially comprising an
  uniform prior. The chains for trigonometric parallax appear to be
  wider due to the narrower allowed parameter space because the
  trigonometric parallax is heavily constrained by that of $\varepsilon$
  Indi A. The same chains are plotted using the same colors for all
  three parameters. Convergence is not evident before 1.4 million
  steps. We conservatively use only the last 100000 steps in inferring
  results.
\label{fig:chains}}
\end{figure*}

\begin{figure}
   \includegraphics[scale=0.6]{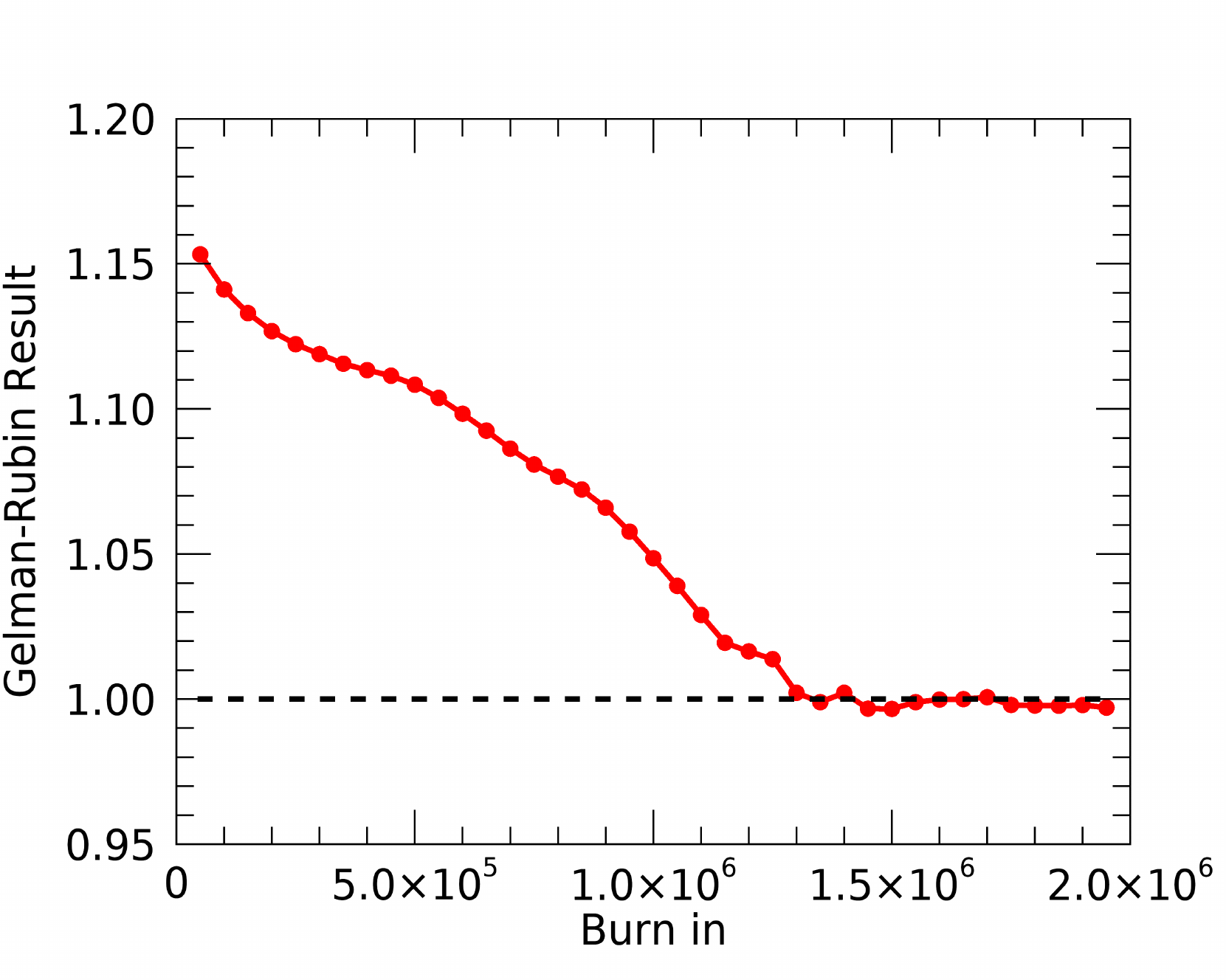}
   \caption{\scriptsize Results of the Gelman-Rubin test for chain convergence.
     The plot indicates the degree of convergence after the number of steps indicated
     in the horizontal axis have been removed as the ``burn in'' phase. Each
     dot indicates an increment of 50000 steps. Results close to 1.00 indicate
     convergence, which is reached after approximately 1.4 million steps.
     The results of the test agree well with the chain plots in Figure \ref{fig:chains}
   \label{fig:GelmanRubin}}
\end{figure}

\begin{figure*}
   \includegraphics[scale=0.25]{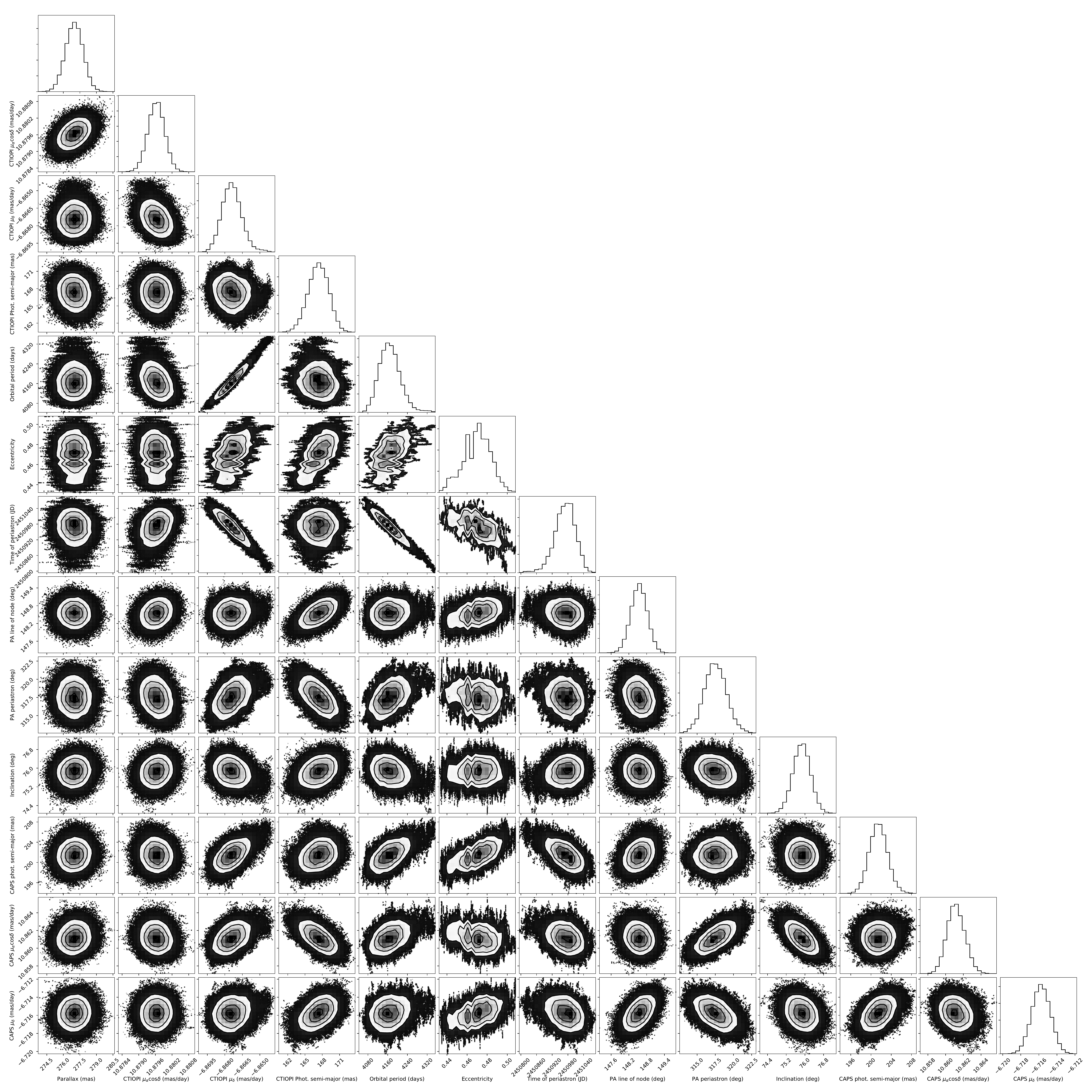}
   \caption{\scriptsize Correlation plots for the 13 astrometric parameters. See
       text for discussion.
    A high resolution version of this Figure is available as an online supplement.
   \label{fig:corner}}
\end{figure*}

\subsection{Dynamical Masses \label{subsec:masses}}

\begin{figure*}[h!]
  \includegraphics[scale=0.6]{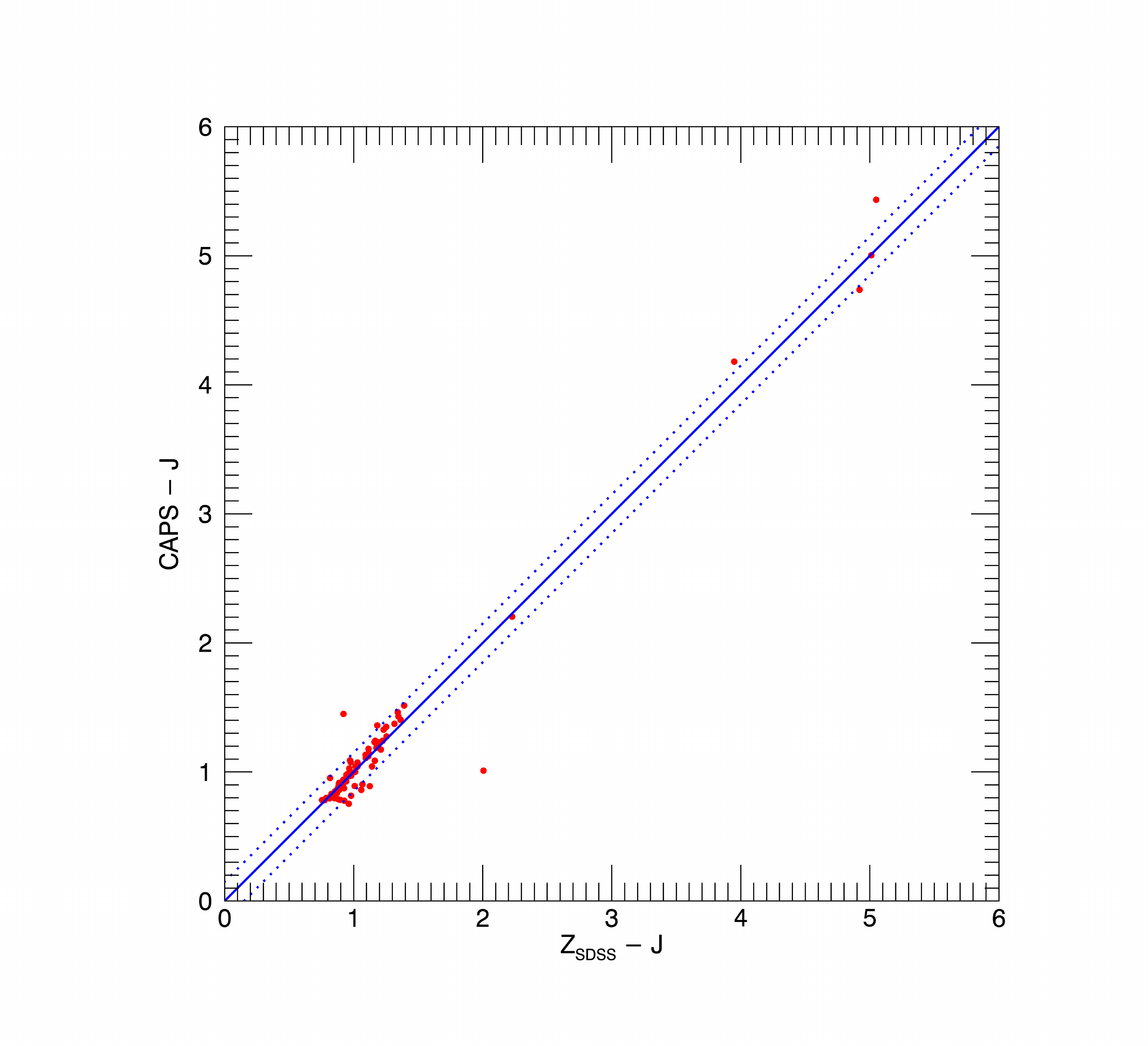}
  \caption{\scriptsize Comparison of the CAPSCam system response
    function to the SDSS $z$ band for stars in the field of LHS
    495. The red dots indicate field objects for which photometry was
    done on both systems.  The solid blue line indicates a 1 to 1
    relation. The dotted lines are the 1$\sigma$ uncertainties of 0.1
    magnitudes about the fit. The $z-J$ colors of the B and C
    components are 2.87 and 3.57, respectively.  \label{fig:zcomp}}
\end{figure*}

To obtain dynamical masses from the photocenter's orbit we followed
the method described in \citet{vandeKamp1968} and
\citet{McCarthyetal1991}. We summarize the formalism here while
providing a detailed example of dynamical mass
determination in Appendix \ref{appendixb}. At any given epoch define
$\rho$ as the magnitude of the photocenter's displacement about the
barycenter and define $p$ as the projected separation between
components B and C. The constant scaling factor from the photocenter's
orbit around the barycenter to the relative orbit of component C
around B is then $p/\rho$ and can be measured at any epoch for which a
resolved image exists. Along with Kepler's Third Law, this relation
yields the system's total mass without regard to the flux ratio of
both components in the unresolved observations.  To obtain individual
masses, define $\mathcal{M}$ as the fractional mass of the C
component:
\begin{equation}
  \mathcal{M} = \frac{M_C}{M_B + M_C}.
\end{equation}
Likewise, define $\mathcal{F}$ as the fractional flux of the C
component in the band used to map the photocenter's
displacement\footnote{The quantities that we denote as
    $\mathcal{M}$ and $\mathcal{F}$ have traditionally been called $B$
    and $\beta$, respectively. We use a different notation to avoid
    confusion with the system's B component.}. The mass ratio is then
found by setting
\begin{equation}
  \frac{p}{\rho} = \frac{1}{\mathcal{M} - \mathcal{F}} ,
\end{equation}
It is therefore necessary to know the flux ratio of the B and C
components in one of the bands used to map the photocenter's
orbit. The SDSS $z$ band very nearly approximates the CAPSCam overall
bandpass \citep{Bossetal2009}.  For this purpose we used the
$z$ band flux ratio from \citet{Kingetal2010} ($F_C/F_B = 0.259 \pm
0.002$).  We validated the photometric equivalency assumption by
comparing $z - J$ colors obtained with the SDSS $z$ band to colors
obtained with CAPSCam for the field of LHS 495, which is very crowded
and observed by both surveys.  Figure \ref{fig:zcomp} shows the
comparison.  While there are only a few very red objects the one to
one relation is clear in the range including the $z-J$ colors of the B
and C components: 2.87 and 3.57, respectively.  We therefore adopt
$\Delta\text{CAPS} \approx \Delta\text{z}$ and add an additional
uncertainty of 0.1 magnitudes to the $F_C/F_B$ flux ratio. We note
that any uncertainty depending on the flux ratio will be propagated to
the mass ratio of the components but will have no effect on the total
mass of the BC system.

\begin{figure*}[h!]
  \includegraphics[scale=0.6]{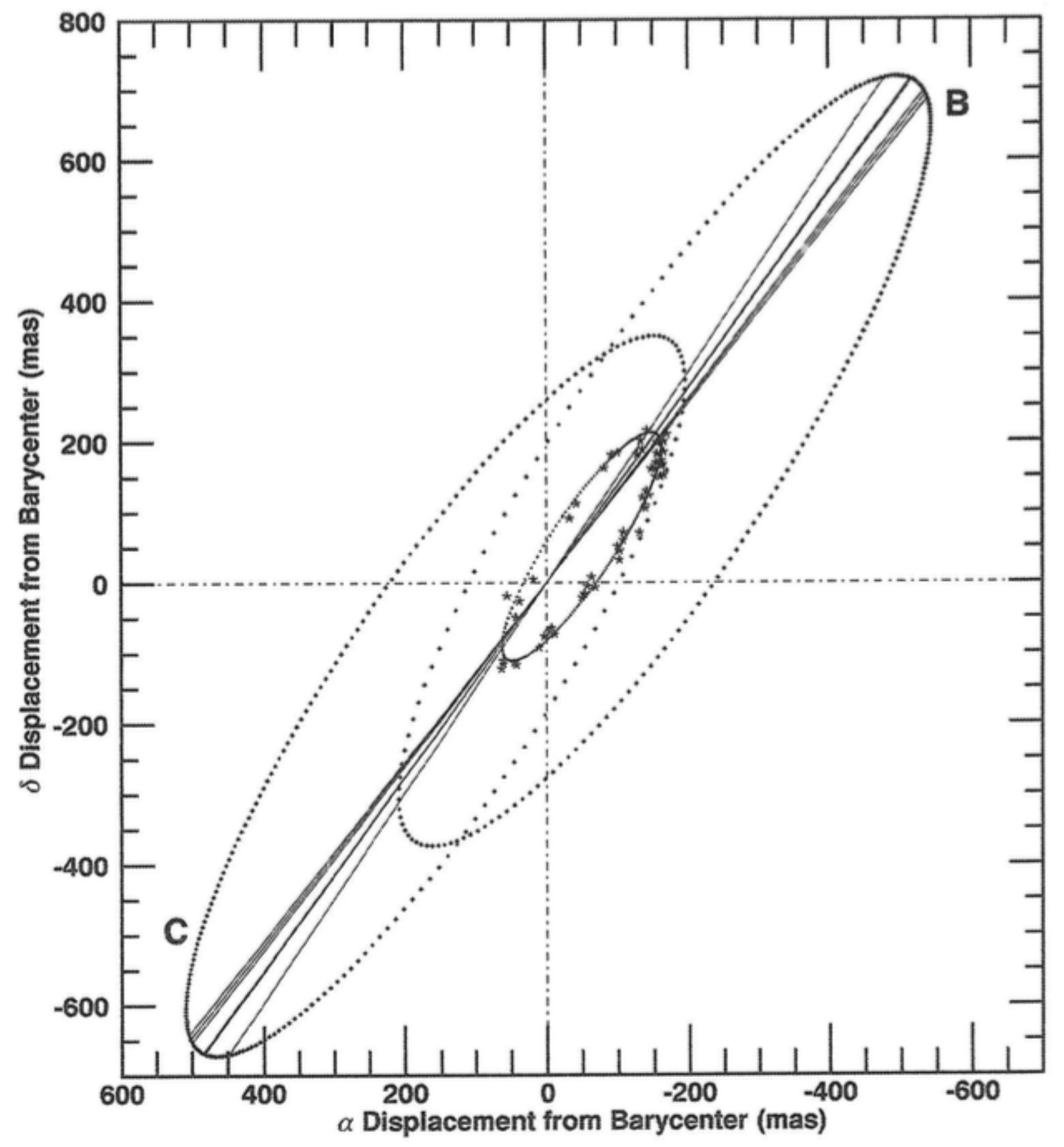}
  \caption{\scriptsize The projected barycentric orbits of $\varepsilon$ Indi B
    and C are plotted along with the photocenter's orbit. The solid
    lines connecting the orbits through the barycenter show the
    separations at the six epochs of high resolution imaging. The observed
    displacements of the photocenter are plotted as asterisks,
    omitting error bars for clarity. We refer the reader to Figure \ref{fig:orbitcontour}
    for a more detailed description of the central portion of this
    Figure. The uncertainty contours shown in Figure \ref{fig:orbitcontour} can be scaled
    linearly to the orbits of the two components. \label{fig:orbits}}
\end{figure*}

Table \ref{tab:masses} lists the dynamical masses we obtained from the
weighted mean of six epochs of high resolution imaging, as well as the
semi-major axes for the orbits of the B and C components about the BC
barycenter. The semi-major axis of the relative orbit of the C
component around the B component, the quantity that is used in solving
Kepler's Third Law, is 2.61$\pm$0.03\,a.u. The total system mass is
0.138$\pm$0.0010\,M$_{\sun}$ or 144.49$\pm$1.06\,M$_{Jup}$. As
previously discussed, this relative semi-major axis and the total
system mass are independent of any photometric flux assumption. The
adopted best values for the individual dynamical masses are
75.0$\pm$0.82\,$M_J$ for the B component and 70.1$\pm$0.68\,$M_J$ for
the C component. Figure \ref{fig:orbits} shows the barycentric orbits
of the individual components along with the photocenter's orbit and
the separations measured in the high resolution images.

\begin{longrotatetable}
  \begin{deluxetable}{cccccccccccccccccc}
\tabletypesize{\scriptsize}
\tablecaption{VLT/NACO Observations, Semi-major Axes, and Dynamical Masses \label{tab:masses}}
\tablehead{ \colhead{Night}     &
            \colhead{JD}         &
            \colhead{Sep.}   &
            \colhead{$\sigma_{sep.}$} &
            \colhead{P.A.\tablenotemark{a}} &
            \colhead{$\sigma_{P.A.}$}   &
            \colhead{$a_B$\tablenotemark{b}} &
            \colhead{$\sigma(a_B)$ }   &
            \colhead{$a_C$\tablenotemark{b}} &
            \colhead{$\sigma(a_C)$ }   &
            \colhead{$M_B$}  &
            \colhead{$\sigma(M_B)$}   &
            \colhead{$M_B$}  &
            \colhead{$\sigma(M_B)$}   &
            \colhead{$M_C$}  &
            \colhead{$\sigma(M_C)$}   &
            \colhead{$M_C$}  &
            \colhead{$\sigma(M_C)$}   \\
            \colhead{      }  &
            \colhead{      }  &
            \colhead{mas}   &
            \colhead{mas}   &
            \colhead{deg.}  &            
            \colhead{deg.}  &            
            \colhead{au}    &
            \colhead{au}    &
            \colhead{au}    &
            \colhead{au}    &
            \colhead{$M_{\sun}$}  &
            \colhead{$M_{\sun}$}  &
            \colhead{$M_{Jup}$}  &
            \colhead{$M_{Jup}$}  &
            \colhead{$M_{\sun}$}  &
            \colhead{$M_{\sun}$}  &
            \colhead{$M_{Jup}$}  &
            \colhead{$M_{Jup}$}    }
\startdata            
    2004-11-11  &   2453323.5 & 894.2 & 2.7 & 140.7 & 0.16 &  1.269 & 0.044 & 1.360 &  0.054  &  0.0722 &  0.0025 &   75.7  &   2.6 &   0.0673 &  0.0018 &   70.5  &   1.9    \\
    2005-06-04  &   2453525.5 & 927.7 & 2.3 & 142.5 & 0.10 &  1.266 & 0.043 & 1.350 &  0.048  &  0.0710 &  0.0019 &   74.4  &   2.0 &   0.0666 &  0.0016 &   69.7  &   1.7    \\
    2005-07-06  &   2453557.5 & 932.3 & 1.1 & 142.6 & 0.10 &  1.267 & 0.043 & 1.352 &  0.047  &  0.0713 &  0.0019 &   74.8  &   2.0 &   0.0667 &  0.0016 &   69.9  &   1.7    \\
    2005-08-06  &   2453588.5 & 934.8 & 2.3 & 142.9 & 0.07 &  1.266 & 0.043 & 1.352 &  0.047  &  0.0713 &  0.0018 &   74.7  &   1.9 &   0.0667 &  0.0015 &   69.9  &   1.6    \\
    2005-12-17  &   2453721.5 & 940.4 & 0.6 & 144.1 & 0.10 &  1.268 & 0.043 & 1.357 &  0.046  &  0.0720 &  0.0018 &   75.5  &   1.9 &   0.0672 &  0.0015 &   70.4  &   1.6    \\
    2005-12-31  &   2453735.5 & 940.0 & 2.2 & 144.1 & 0.18 &  1.268 & 0.043 & 1.358 &  0.046  &  0.0720 &  0.0018 &   75.5  &   1.9 &   0.0672 &  0.0016 &   70.4  &   1.7    \\
{\bf Weighted Means}  & \nodata & \nodata   & \nodata & \nodata & \nodata &  {\bf 1.267} & {\bf 0.018} & {\bf 1.355} & {\bf 0.0195} &  {\bf 0.0716} & {\bf 0.0008} & {\bf 75.0}  & {\bf 0.82} & {\bf 0.0669} & {\bf 0.00064} & {\bf 70.1}  & {\bf 0.68}    \\
\enddata
\tablenotetext{a}{Position angle of C relative to B measured E of N. Differs by $\pm$180$^{\circ}$  from photocenter's orbit.}
\tablenotetext{b}{Derived semi-major axis of component's orbit about the BC system's barycenter using the ratio of the photocentric offset to the observed separation at that epoch.}
  \end{deluxetable}
\end{longrotatetable}

\subsection{Statistical Tests, Convergence, and Systematic Errors \label{subsec:statistics}}
Figure \ref{fig:chains} shows the evolution and convergence of Markov
chains for the three astrometric parameters used in dynamical mass
determination: trigonometric parallax, semi-major axis, and orbital
period. Based on the long burn in phase of several chains, we
conservatively choose to use only the last 100000 samples. We formally
verified the convergence of the 52 Markov chains by applying the
Gelman-Rubin statistical test for convergence
\citep{GelmanAndRubin1992}\footnote{Implemented in IDL by Simon Vaughan,
  University of Leicester, https://www.star.le.ac.uk/$\sim$sav2/idl/rhat.pro}.
Figure \ref{fig:GelmanRubin} shows the
results as a function of the length of the chain's burn in phase. The
test measures the extent to which all 52 chains have converged to a
stable result, indicated by a value approximating 1.00. The test
confirms what is seen graphically in Figure \ref{fig:chains} $-$ that
convergence is obtained after about 1.4 million steps and the chains
have completely stabilized in the last 100000 steps, from which we
draw the astrometric parameters.

Figure \ref{fig:corner} shows the correlation plots for all 13
astrometric parameters.  Most parameter combinations show low or no
correlation with well defined central values.  Combinations of the
temporal parameters of proper motions, time of periastron passage, and
orbital period show very high correlation, as is to be expected of
parameters that determine the photocenter's displacement as a function
of the same critical domain. Further, most parameters show mild
correlation with the proper motion parameters, most likely due to the
fact that proper motion is by far the dominant source of the system's
displacement.

We save a discuss of the MCMC's acceptance fraction for Appendix \ref{appendixa}.

  \subsubsection{Error Analysis of the Mass Derivation \label{subsubsec:masserror}}
  The dynamical masses listed in Table \ref{tab:masses} include
  Gaussian uncertainties propagated via Monte Carlo, which is
  appropriate for independent random uncertainties.  We now examine
  the possibility of systematic errors in the dynamical masses.

  From Appendix \ref{appendixb}, the quantities necessary for the
  dynamical mass calculation are the physical semi-major axis of the
  relative orbit($a$), the orbital period ($P$), the orbital
  separation at a given epoch ($p$), and the displacement of the
  photocenter calculated at the same epoch ($\rho$). The quantities
  $a$ and $p$ are functions of the photocentric semi-major axis
  ($\alpha$), the trigonometric parallax ($\Pi$), and the pixel scale
  used in the VLT/NACO observations. Given the excellent agreement
  between the parallax we obtain (276.88$\pm$0.81\,mas) and the
  Hipparcos parallax for the A component (276.06$\pm$0.28\,mas) we can
  rule out systematic errors in our trigonometric parallax. The pixel
  scale of the NACO detector was examined in detail by
  \citet{Ginskietal2014}, who find a mean value of
  13.233$\pm$0.012\,mas/pixel based on five globular cluster
  calibrations between 2008-06-14 and 2012-03-03.  This value is
  within 0.28 percent of the pixel scale reported by the observatory
  and used in our calculations, 13.270\,mas/pixel. propagating this
  offset leads to a 0.841 percent {\it increase} in the dynamical
  masses: 0.63\,$M_{Jup}$ for $\varepsilon$ Indi B and 0.59\,$M_{Jup}$
  for the C component. These offsets are within the uncertainties of
  our reported masses, 75.0$\pm$0.82\,$M_{Jup}$ and
  70.1$\pm$0.68\,$M_{Jup}$ for the B and C components, respectively.
  The uncertainty in pixel scale, caused by any systematics or
  temporal drift, is therefore not a significant source of
  error and is most likely contributing to the uncertainties we
  already adopt.

  The displacement of the photocenter at the epoch of AO observation 
  $\rho$ and the semi-major axes of the relative physical orbit $a$
  are also functions of the photocentric orbit's semi-major axis, $\alpha$.
  This last quantity is inferred by the MCMC, which raises the question
  of wether or not all parameters inferred by the MCMC are correct and not
  contributing systematic error to the other parameters. The result of the Gelman-Rubin
  test (Figure \ref{fig:GelmanRubin}) strongly suggests that all parameters
  have converged to their true values. As a further test we examine
  parameter correlations and wether or not they could be offsetting each other.
  For the purposes of mass calculation the relevant MCMC
  inferred parameters are $\alpha$, $P$, and $\Pi$. In principle,
  these values could
  contain systematic errors if the sources of motion, the trigonometric
  parallax, the proper motion, and the orbital motion, are not
  well separated. Figure \ref{fig:corner} shows the correlation plots
  for all parameters. The parameters $\alpha$ and $P$ show slight correlation
  to each other and $\alpha$ is also correlated to the declination component
  of proper motion. Correlation is not necessarily a sign
  of systematic error, but it raises the possibility that those parameters
  could be contributing to each other in an erroneous manner. 
  To test this, we introduced 4$\sigma$
  systematic errors to the declination component of proper motion,
   $\alpha$, and $P$ and held each one of those parameters fixed
  while testing the MCMC for convergence. In all three cases the
  MCMC results did not converge, resulting in multi-modal distributions
  as opposed to the well-defined gaussian probability density functions
  shown in Figures \ref{fig:pdf1}, \ref{fig:pdf2}, and \ref{fig:pdf3}.
  We therefore conclude that the probability of erroneous
  interference amongst the sources of motion in our solution
  is very low.

\section{Discussion}\label{sec:discussion}

$\varepsilon$ Indi B and C are unique in being the only T dwarfs with
known dynamical masses that approach the theoretical hydrogen burning
minimal mass limit. \citet{Konopackyetal2010} obtained a total system
dynamical mass of 62\,$M_{Jup}$ for the T5.5+T5.5 binary 2MASS
J10210969$-$304197, from which we infer individual masses of
approximately 31\,$M_{Jup}$. \citet{DupuyandLiu2017} report individual
component dynamical masses for six T dwarf ranging from 31\,$M_{Jup}$
to 55\,$M_{Jup}$. Most recently, \citet{Bowleretal2018} obtained a
dynamical mass of 42$^{+19}_{-7}$\,$M_{Jup}$ for the late T dwarf GJ
758 B, all of which are firmly in the substellar mass range.
  While the determination of these dynamical masses are valuable contributions,
  it is difficult to constrain evolutionary models with masses that are firmly
  in the substellar domain because of the large degeneracy between mass and age
  in the brown dwarf cooling track. As an example, \citet{Burrowsetal2001}
  predict that a mid T dwarfs could range in mass from about 20\,$M_{Jup}$ to
  about 60\,$M_{Jup}$ if these their ages are 500\,Myr and 10\,Gyr, respectively.
  Both ages are possible within the general galactic disk population.
  It is therefore not surprising that the components of 2MASS J10210969$-$304197
  and $\varepsilon$ Indi C have approximately the same spectral type and vastly
  different masses because they probably have very different ages.
  Because there are no reliable age indicators for isolated older substellar
  objects there is little we can learn from them regarding substellar cooling rates.
  In contrast, objects with masses close to the stellar-substellar limit
  allow us to test a boundary value of the theory of substellar structure and evolution.

 There is broad consensus that T dwarfs are substellar objects
\citep[e.g.][]{Kirkpatrick2005, Burrowsetal2001}.  Table
\ref{tab:hbmm} lists the most widely used evolutionary models for
substellar objects and the temperatures at which they predict the
stellar-substellar boundary.  \citet{Kingetal2010} infer effective
temperatures in the range of 1,300\,K to 1,340\,K and 880\,K to 940\,K
for $\varepsilon$ Indi B and C, respectively. These temperatures are
in good agreement to those inferred by \citet{Filippazzoetal2015} for
field-aged objects of spectral types T1 and T6: 1,200\,K and
900\,K. We see therefore that virtually the entirety of our
theoretical understanding of T dwarfs would be significantly
 off if $\varepsilon$ Indi B and C were stellar
objects. As we are about to discuss, problems with the theory do
exist. However, a scenario that makes $\varepsilon$ Indi B and C, or
even only the B component, a star is extremely unlikely; all models
listed in Table \ref{tab:hbmm} would have to be
over-predicting the temperature of the stellar-substellar
  boundary by several hundred Kelvins. And yet the masses we obtained
are remarkably high given our current understanding of substellar
structure and evolution. There is strong evidence from photometry,
spectroscopy, and adaptive optics imaging that the B and C components
have different luminosities and therefore must have different masses
given their common age (Sections \ref{sec:intro} and
\ref{subsec:masses}).  However, even if we disregard flux ratio,
dividing the total system mass by two means that the most massive
component cannot have a mass under 72.5 M$_{Jup}$. We note also that
while the $\varepsilon$ Indi system has slightly sub-solar metallicity
(Table \ref{tab:epsind}) stars with [Fe/H]$\approx -$0.1 are common in
the solar neighborhood and should be included when considering the
(sub)stellar population as a whole. \citep[e.g.][]{Hinkeletal2014}.

\begin{figure*}
  \plottwo{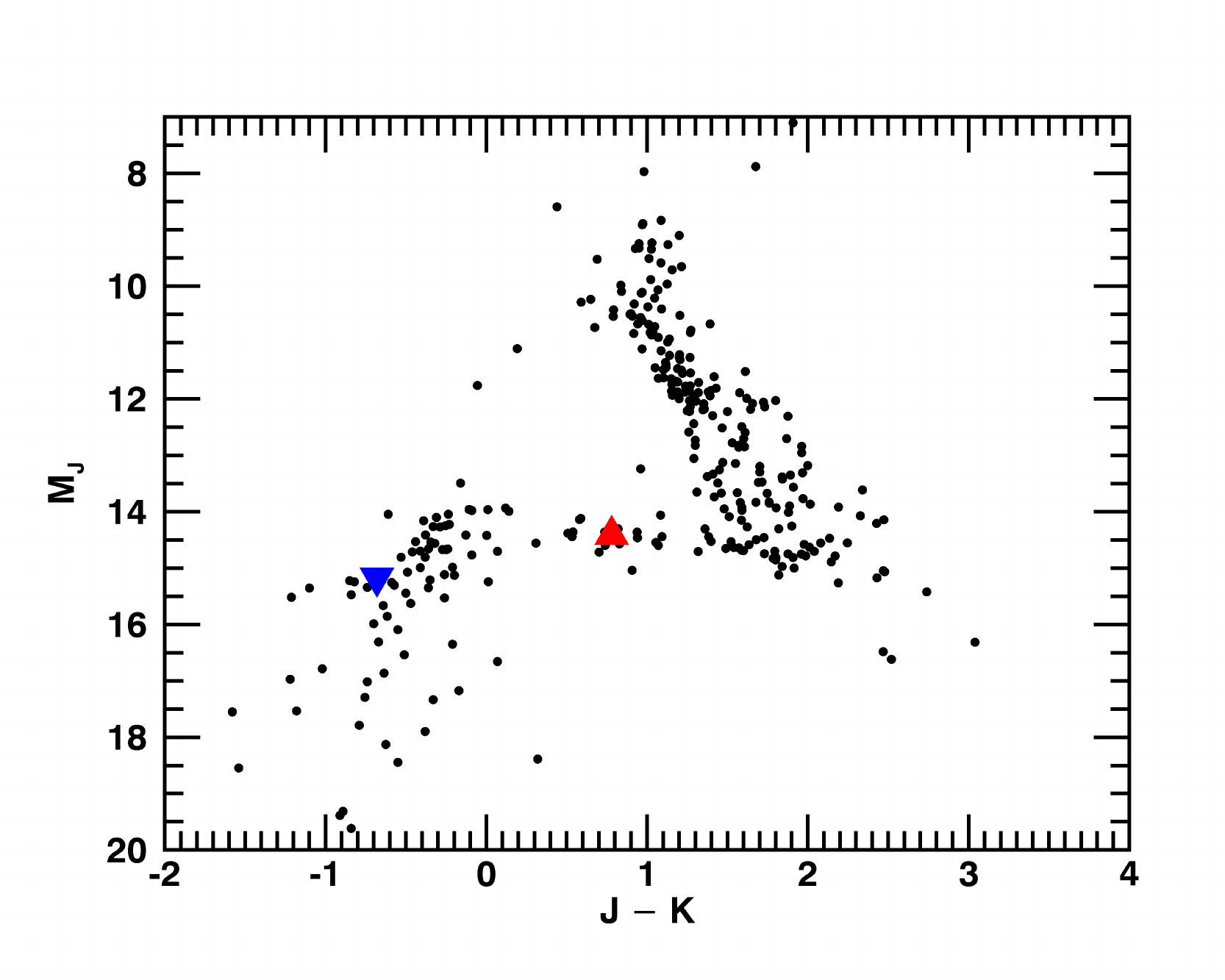}{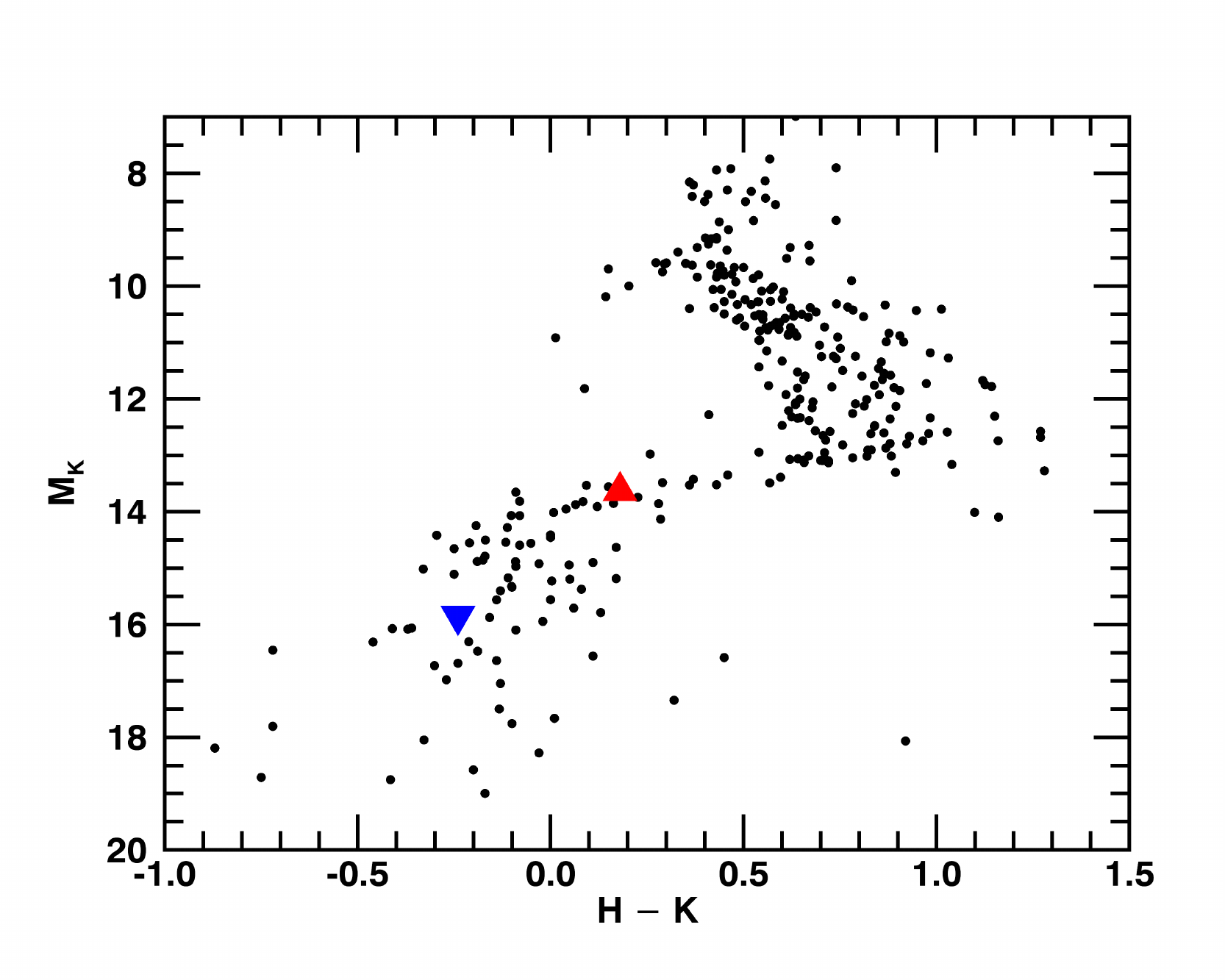}
  \caption{\scriptsize Color-magnitude plots showing e Indi B plotted
    as a red triangle and C as an inverted blue triangle. While the C
    component appears to lie slightly blueward of the center of the
    sequence, neither component can be distinguished from the general
    population. While some objects are clearly deviant from the
    sequence, in the interest of completeness we did not exclude any
    data from T. Dupuy's database. \label{fig:cmd}}
\end{figure*}

Figure \ref{fig:cmd} shows two color-magnitude diagrams based on the
data from the Database of Ultracool Parallaxes maintained by Trent
Dupuy\footnote{currently hosted at
  http://www.as.utexas.edu/tdupuy/plx/Database\_of\_Ultracool\_Parallaxes.html}
\citep{DupuyandLiu2012, DupuyandKraus2013}.  While the C component
appears to be slightly blue, neither component stands out from the
general field population. Any correct theoretical framework must allow
such massive objects to be substellar and also provide an
appropriately fast cooling rate so that they reach the T spectral type
in a time that must be smaller than the upper bound on the system's
age, the age of the Galaxy.

Table \ref{tab:hbmm} lists several studies regarding the properties of
the stellar-substellar boundary from both theoretical and
observational perspectives. All theoretical treatments agree that as a
general trend, higher opacity moves the end of the stellar main
sequence downward in mass, luminosity, and effective temperature. The
overall opacity driving the trend is a complex function of metallicity
and atmospheric parameters such as the presence of silicate grains and
their sedimentation rate. The result is a wide range of predictions
for the fundamental parameters of the smallest and least
  massive possible stellar objects. At present, the models that
address the effects of different metallicities and atmospheric
conditions tend to consider extreme values, so it is difficult to
ascertain what the models would predict in the case of $\varepsilon$
Indi's slightly sub-solar metallicity.  We can gain some insight from
the fact that the models listed in Table \ref{tab:hbmm} have adopted
different values for the zero point solar abundances and the offsets
among those different values provide a limited range of metallicities
for comparison.

\begin{deluxetable*}{lccccccl}[h!]
\tabletypesize{\scriptsize}
\tablecaption{Summary of Predictions for the Stellar-Substellar Boundary\tablenotemark{a}\label{tab:hbmm}}
\tablehead{
            \colhead{Study}    &
            \colhead{Type}     &
            \colhead{H. Burning}   &
            \colhead{H. Burning}   &            
            \colhead{H. Burning}   &
            \colhead{Metallicity\tablenotemark{b}}   &
            \colhead{Min. Stellar}    &
            \colhead{Atmospheric}     \\
            \colhead{      }       &
            \colhead{of Study}     &
            \colhead{Mass ($M_{Jup}$)}      &
            \colhead{$T_{eff}$ (K)}    &
            \colhead{Log($L/L_{\sun}$)}  &
            \colhead{($Z/Z_{\sun}$)}    &
            \colhead{Radius ($R/R_{Jup}$)} &
            \colhead{Properties}
            }
\startdata
\citet{Burrowsetal1993,Burrowsetal1997}  &    Model        &      80.4     &  1747    &   -4.21    &   1.28    &     0.84    &  gray with grains                     \\
\citet{Burrowsetal1993}                  &    Model        &      98.5     &  3630    &   -2.90    &   0.00    &     0.89    &  metal free\tablenotemark{c}          \\
\citet{Baraffeetal1998}                  &    Model        &      75.4     &  1700    &   -4.26    &   1.28    &     0.84    &  non-gray without grains              \\
\citet{Chabrieretal2000}                 &    Model        &      73.3     &  1550    &   -4.42    &   1.28    &     0.86    &  ``DUSTY'' grains do not settle         \\
\citet{Burrowsetal2001}                  &    Model        & 73.3 to 96.4  &  \nodata &   \nodata  &   various &   \nodata   &  discussion of various models         \\
\citet{Baraffeetal2003}                  &    Model        &      75.4     &  1560    &   -4.47    &   1.28    &     0.81    &  ``COND'' clear and metal depleted\tablenotemark{d} \\
\citet{SaumonandMarley2008}              &    Model        &      78.6     &  1910    &   -4.00    &   0.87    &     0.89    &  cloudless                            \\
\citet{SaumonandMarley2008}              &    Model        &      73.3     &  1550    &   -4.36    &   0.87    &     0.91    &  cloudy, $f_{sed} = 2$                     \\
\citet{Baraffeetal2015}                  &    Model        &      73.3     &  1626    &   -4.30    &   1.00    &     0.89    &  ``BT-Settl'' cloud model               \\
\citet{Dieterichetal2014}                &    HR Survey    &   \nodata     &  2075    &   -3.90    &   Field   &     0.86    &   \nodata                                    \\
\citet{DupuyandLiu2017}                  &    Mass Survey  &      70       &  \nodata &   \nodata  &   Field   &     \nodata &   \nodata                             \\
\hline                                                                                                                                                               
{\bf $\varepsilon$ Indi B}               &    Mass         & {\bf 75.0$\pm$0.82}  &  1320\tablenotemark{e}  &   -4.70\tablenotemark{e}   &    0.74   &   0.83 &   \nodata                             \\
{\bf $\varepsilon$ Indi C}               &    Mass         & {\bf 70.1$\pm$0.68}  &  910\tablenotemark{e}   &   -5.23\tablenotemark{e}   &    0.74   &   0.85 &   \nodata                             \\  
\enddata
\tablenotetext{a}{Adapted and updated from Table 8 of \citet{Dieterichetal2014}}
\tablenotetext{b}{We adopt the solar metallicities of \citet{Caffauetal2011} as the ``true'' solar value.
  With the exception of the zero
  metallicity case of \citet{Burrowsetal1993} all models were meant as
  solar metallicity when they were published. Here we scaled their
  metallicities to reflect the new values of \citet{Caffauetal2011}. See
  \citet{Allardetal2013} for a discussion of recent revisions to solar
  abundances.}
\tablenotetext{c}{An artificial case meant to illustrate the significance of metallicity in determining the parameters
    of the stellar-substellar boundary.}
\tablenotetext{d}{In this case metals are sequestered in grains that settle below the photosphere.}
\tablenotetext{e}{\citet{Kingetal2010}}
\end{deluxetable*}

Given the uncertainties on the dynamical mass of $\varepsilon$ Indi B
(75.0$\pm$0.82\,$M_{Jup}$) none of the models listed in Table \ref{tab:hbmm}
can be strictly ruled out as far as predicting its substellar
nature. However, \citet{Chabrieretal2000}, the cloudy version of
\citet{SaumonandMarley2008}, and \citet{Baraffeetal2015}, the only
evolutionary model in that family to use updated metallicities
\citep{Caffauetal2011} and an adaptive cloud model
\citep{Allardetal2013}, do so only marginally and predict masses
$\sim$2$\sigma$ away from our best values.

Our results are not compatible with the possibility of
the stellar-substellar mass boundary being at 70\,$M_{Jup}$, as
claimed by \citet{DupuyandLiu2017}. We note, however, that their own data
\citep[][Table 11 and Figure 3]{DupuyandLiu2017} are amenable to broad
interpretation and that the mean of the
uncertainty of their dynamical masses that fall within the 70\,$M_{Jup}$
range is 5.6\,$M_{Jup}$.
Finally, \citet{Dieterichetal2014}
examined luminosity, temperature, and radius trends in the early L
dwarf range and did not address the question of the minimum stellar
mass directly. The claims in \citet{Dieterichetal2014} therefore
cannot be directly tested by the dynamical masses we report here.

Assuming that a given model predicts that such massive objects would
be substellar, another relevant question is wether or not the model
can provide a fast enough cooling rate to make $\varepsilon$ Indi B
and C reach T dwarf
temperatures.\citet{Filippazzoetal2015} establish
  luminosities (Log($L/L_{\sun}$)) of $-$4.6 and $-$5.0 for T1.5 and
  T6 dwarfs respectively, and effective temperatures of 1,200\,K and
  900\,K for field-aged objects. Most models we consider in Table
\ref{tab:hbmm} cannot make such massive objects
  reach those temperatures in times shorter than
the age of the Galaxy.  Even at an age of 10\,Gyr the
Burrows solar metallicity models
\citep{Burrowsetal1997,Burrowsetal2001} predict the effective
temperature of a 75\,$M_{Jup}$ object to be 200\,K to 400\,K hotter
than 1,200\,K, thus placing $\varepsilon$ Indi B in the mid to late L
spectral type range. The discrepancy for the C component is slightly
smaller, on the order of 150\,K. The same models can accommodate the
luminosities and temperatures of both components at 10\,Gyr if the
opacity is decreased by setting [Fe/H] = $-$1.0; however, that is a
factor of 7.4 less than the observed metallicity of $\varepsilon$ Indi
A ([Fe/H] = $-$0.13, see Table \ref{tab:epsind}). See Figures 1, 4, 5, and 8
of \citet{Burrowsetal2001} for graphical representations of these models.

The cloudy version
  of the \citet{SaumonandMarley2008} and the ``DUSTY'' models of
  \citet{Chabrieretal2000} are similar to each other at ages greater
  than 4\,Gyr. They predict temperatures about 400 K above the
  temperatures from the \citet{Filippazzoetal2015} field sequence. See
    Figure 3 of \citet{SaumonandMarley2008} for graphical representations
    of these models.

The ``COND'' models of \citet{Baraffeetal2003} and the cloudless
version of \citet{SaumonandMarley2008} deliberately simulate less
opaque atmospheres to investigate the possibility that atmospheric
silicate grains either do not form or settle quickly below the
photosphere. These models predict considerably faster cooling rates,
and can reach the temperatures of $\varepsilon$ Indi B and C at ages
$\lesssim$ 10 Gyr if their masses are slightly beyond the lower bounds in
the 1$\sigma$ uncertainties we obtained. The $\varepsilon$ Indi system would
then have to be very old to fit these isochrones. See
    Figure 2 of \citet{SaumonandMarley2008} for graphical representations
    of these models. The K5V A component
has been associated with the small moving group of the same name \citep{Eggen1958}, and
age estimates for that moving group range from 5\,Gyr to 6.2\,Gyr
\citep{Kingetal2010, SoubiranandGirard2005, Cannon1970}.
However the proposed moving group is small, with only 15 stars identified
by \citet{Eggen1958}, and \citet{KovacsAndFoy1978} cast doubts into its membership
and existence. \citet{Lachaumeetal1999} note that while the
chromospheric activity of $\varepsilon$ Indi A indicates an age of 1
to 2.7\,Gyr the system's Galactic kinematics are consistent with a much
older age of $\gtrsim$7.4\,Gyr.  It is therefore plausible that the
system is old, but as is evident from this discussion, the ages of
low mass main sequence stars are notoriously difficult to determine.
A greater problem may be that while cool brown dwarfs that have passed
the L-T transition are thought to have clear atmospheres, the spectra
of L dwarfs are well replicated by models that include some amount of
silicate grains \citep[e.g.,][]{Allardetal2013}. The conditions
assumed in these clear models may therefore not be a realistic
representation of a T dwarf's cooling history. Nevertheless, we note
that the ``COND'' models of \citet{Baraffeetal2003} and the cloudless
models of \citet{SaumonandMarley2008} come considerably closer to
matching the parameters of $\varepsilon$ Indi B and C than the more
opaque models also listed in Table \ref{tab:hbmm}.

\section{Conclusions\label{sec:conclusions}}
We inferred the dynamical masses of $\varepsilon$ Indi B and C
using unresolved photocentric astrometric data,
resolved adaptive optics images, and Markov Chain Monte Carlo
techniques. The dynamical masses we obtained, 75.0$\pm$0.82\,$M_{Jup}$ and
70.1$\pm$0.68\,$M_{Jup}$ for the B and C components, respectively, are
surprisingly high and challenge our understanding of the
stellar-substellar mass boundary.

Our analysis highlights the strengths and weaknesses of different
approaches to understanding the structure and evolution of two old and
massive brown dwarfs. It is clear that the current models under-predict
the upper mass limit and/or the necessary cooling rates for $\varepsilon$
Indi B and C, with less opaque models coming closer to replicating the
observed parameters. The system's slight negative departure from solar
metallicity is of the same order as the differences among the solar
abundances chosen by different models listed in Table
\ref{tab:hbmm}. The lack of a clear trend linking metallicities to
fundamental parameters in Table \ref{tab:hbmm} as well as the lack of
a clear displacement from the field sequence in color-magnitude
diagrams (Figure \ref{fig:cmd}) suggest that small changes in
metallicity are likely not a dominant factor in determining the
structure and evolution of brown dwarfs. This result supports the
theoretical argument to the same effect first proposed by
\citet{Burrowsetal2001}.

\citet{Dieterichetal2014} examined the fundamental parameters of the
nearby field L dwarf population and found that the models likely
under-predict the luminosity and effective temperature of objects at
the stellar-substellar boundary. While that study did not directly
address mass, the most fundamental of all (sub)stellar parameters,
they noted that if the higher than expected temperatures and
luminosities were due to models overestimating opacities, then a higher
limit for substellar masses should also be expected. The high
dynamical masses of $\varepsilon$ Indi B and C support that
explanation.

Finally, we note that we now have two well characterized brown dwarfs
with precise dynamical masses very close to the stellar-substellar
boundary, spatially resolved spectra and photometry, and well
constrained metallicities from their main sequence primary
component. The thorough modeling of the $\varepsilon$ Indi system
using precise empirical input values would provide considerable insight
about outstanding theoretical issues regarding the
stellar-substellar boundary.

We thank John Gizis and G. Fritz Benedict for helpful discussions
and the anonymous referees, whose insights have helped to greatly improve
the article.  S.B.D. acknowledges support from the National Science Foundation
Astronomy and Astrophysics Postdoctoral Fellowship Program through
grant AST-1400680. We thank the David W. Thompson Family Fund for
support of the CAPSCam astrometric planet search program and the
Carnegie Observatories for continued access to the du Pont
telescope. The development of the CAPSCam camera was supported in part
by NSF grant AST-0352912. The CTIOPI program is made possible through
the SMARTS Consortium and through NSF grants AST-0507711, AST-0908402,
AST-1412026, and AST-1715551.  This research has made use of the
SIMBAD database, operated at CDS, Strasbourg, France. This research
has made use of the services of the ESO Science Archive Facility.
This research has made use of IRAF, which
is distributed by the National Optical Astronomy Observatory,
operated by the Association of Universities for Research in Astronomy
(AURA) under a cooperative agreement with the National Science
Foundation.

We wish to honor the memory of our colleague Sandra Kaiser, an
integral member of the CAPS team who died while this work was being
done. Carnegie is not the same without Sandy.

\appendix

\section{The Markov Chain Monte Carlo Algorithm and its Implementation} \label{appendixa}

The goal of the MCMC is to generate samplings from posterior probability density functions
  that can then be interpreted through Bayesian inference as the probability density functions
  for parameters of interest. We found that solving the astrometric
problem described in Section \ref{sec:model} poses two specific
challenges. First, step sizes must be generated in an adaptive manner
that prevents a single parameter from dominating the overall
evolution of the probability density function. This is particularly
  problematic because the orbit's orientation angles may take on values that minimize
  or maximize the effect of a given direction of motion while probing the parameter space.
Second, a mechanism must exist to cause
any chains stuck in local probability maxima to continue to evolve. We
developed a modification of the Metropolis$-$Hastings MCMC procedure
that specifically addresses these issues in the context of the
astrometric problem. Here we describe the general case assuming a single astrometric
data set. The extension to two or more data sets is not difficult and is discussed
in Section \ref{sec:model}. The IDL suite of codes and detailed documentation
are available for download at
\begin{center}
  github.com/SergeDieterich/MCMC\_SD.
\end{center}
This standard implementation also allows the user to set any parameter
to a fixed number so as to solve a specific subset of the astrometric problem, such as a
resolved visual binary or the trigonometric parallax for a single star. Readers
are encouraged to contact S.B.D. to discuss non-standard uses.

Assuming Gaussian uncertainties in the observed positions of the
photocenter (Table \ref{tab:astro}) then the
probability that a given set of astrometric parameters matches the data
is expressed as
\begin{equation}
  \text{ln}P = -\frac{1}{2}(\chi^{2}_{\alpha} + \chi^{2}_{\delta} + K)
\end{equation}
where $\chi^{2}_{\alpha,\delta}$ refer to equations 10 and 11.
Equation A1 is valid for independent parameters. This is the
  case for displacements and uncertainties in declination and right
  ascension because both telescopes used here are polar mounted,
  meaning their jitter has separate sources for the two orthogonal
  axes. The uncertainties due to atmospheric seeing are also known to
  be isotropic.  The logarithm facilitates numerical computations for
very small probabilities while still preserving an easy way to compare
probability ratios. The term K in equation A1 expresses the
  probability introduced by the choice of prior. However because we
  use constant uniform priors (Section \ref{spider}) K is a constant and cancels out when we
  take the ratio of probabilities between two MCMC steps to evaluate
  wether the chain will advance.

The rule for deciding if a given step advances the Markov chain is a
standard Metropolis$-$Hastings procedure. After a step changes the values of
all astrometric parameters simultaneously the Markov chain will remain
in the new location if the ratio of the new to the previous solution
probability is greater than an uniformly distributed random number in
the range of zero to one and will return to the previous location
otherwise.

The step generator for each parameter is of the form
\begin{equation}
  \text{Step} = A B^{C}\text{ }\frac{\partial(\text{scaling parameter})}{\partial(\text{parameter})} D
\end{equation}
where $A$ and $C$ are randomly discrete $\pm$1, $B$ is a random number between
zero and the user specified ``step multiplier'' parameter, which
governs the scale of the step distribution. $D$ is a fixed step size
for a chosen ``scaling parameter''. The partial derivative ensures
that in the long run of hundreds or more steps all parameters will
contribute more or less equally to the evolution of the Markov chain
towards a solution. The partial derivative is an order of magnitude
calculation and is approximated by measuring the slope of the overall
displacement in the astrometric solution taking the last two steps
into account:
\begin{equation}
  \frac{\partial(\text{ln}P)}{\partial(\text{parameter})} \approx
  \frac{\partial(\sum\limits_{\text{ all epochs}}\Delta(\alpha,\delta))}{\partial(\text{parameter})}.
  \end{equation}
In the work discussed here we chose trigonometric
parallax as the scaling parameter, set $D = 1$ and the step multiplier
equal to 10 so that for each parameter in each step $0.1 \leq B^{C} \leq 10$. That setup
caused all parameters to change the value of the overall solution in a
scale similar to changing the trigonometric parallax by 1\,mas for each
step in the long run, but with $A B^C$ causing considerable variation
in the small scale of $\lesssim$100 steps. We set $C = -1$ at every
five steps for all parameters so as to provide small adjustments in
parameter space that would otherwise be difficult to achieve due to the
simultaneous and independent nature of the parameter steps.

\subsection{Avoiding Local Maxima and Enabling Broad Uniform Priors $-$ the ``Spider'' Mechanism} \label{spider}

Little is known {\it a priori} about the specific configuration of an
unresolved binary system other than very broad constraints that can be
inferred from the design of the observations and the nature of the
data. As examples, a binary system's unresolved nature means that its
projected semi-major axis must be below the telescope's resolution,
and the fact that we see non-linear displacements at periods greater
than one year means that the temporal baseline of the observations is
comparable to the orbital period.  We use only these broad assumptions
in defining the ranges of parameter space to be explored, thus
effectively establishing broad uniform priors. Given the wide
diversity in binary systems and the convoluted nature of the parameter
space we are exploring we believe there is little justification to
assume any other form of prior. It is therefore important that our
MCMC algorithm explore the entirety of this broad parameter space and
lose the dependence on the discrete initial values of each chain. 

One of the difficulties we encountered in using MCMC algorithms was
the prevalence of local maxima in probability space that did not
correspond to the true solution.  This problem persisted even after
employing the adaptive step scaling.  To solve this problem we devised
an additional step scale rule that causes a chain to take very large
jumps in parameter space, probe the new region, and then compare the
relative probabilities of the old and new regions to decide in which
region the chain should continue to evolve. This method is similar to
the ``Snooker'' updater of \citep{terBraakAndVrugt2008}. This mechanism gives the
chains the ability to jump between local probability maxima until they
land in the region of the absolute probability maximum. We call this
condition the ``spider'' mechanism in analogy to a spider that extends
one of its legs to a distant region, probes that locality, and then
decides whether or not to move its entire body there or to retract its
leg.

\begin{figure}[h!]
  \includegraphics[scale=1]{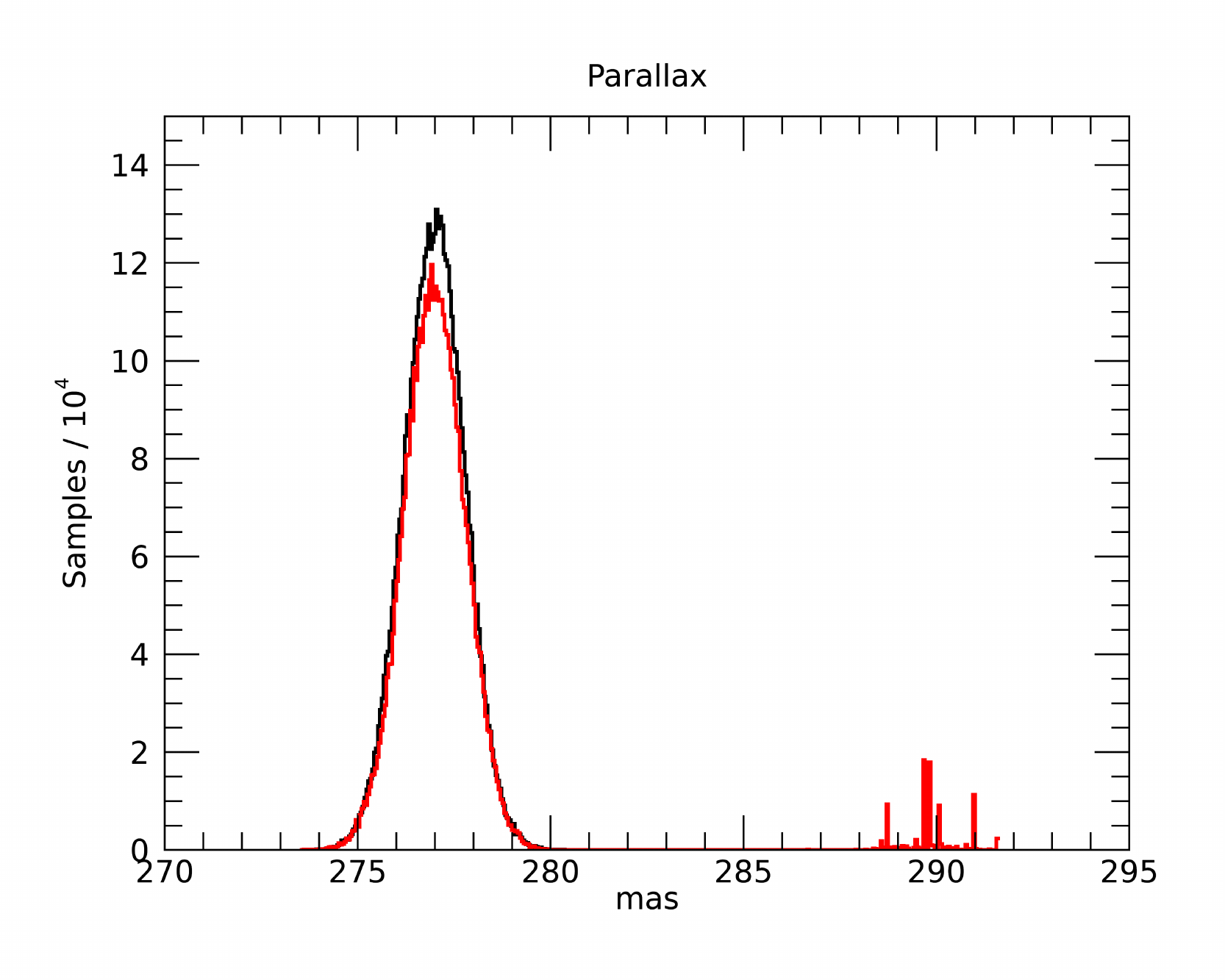}
  \caption{\scriptsize A comparison of the probability density
      function histograms for trigonometric parallax produced while
      employing the spider mechanism (black), and without the spider
      mechanism (red). While the location of the Gaussian peaks are
      very nearly the same the probability density function produced
      without the spider mechanism shows a spurious local maximum
      around 190 mas.  This case is typical of all
      parameters.} \label{fig:spider}
\end{figure}

At every 200 steps one astrometric parameter is randomly selected and
jumps to a new random location within its allowed range. The Markov
chain then evolves normally for 100 steps. If the mean solution
probability of the 100 steps after the jump is less than the mean
probability of the 100 steps before the jump the chain will return to
its pre-jump location, or continue in the new vicinity otherwise.
 This mechanism is most effective in causing large jumps early on
during the MCMC evolution, when all probabilities are likely to be
very low. Once the chains resemble the final probability density
functions most spider jumps are rejected or cause shifts that are
comparable to the normal step process. This ensures that the ergodicity
of our MCMC algorithm is not broken by this additional mechanism
\citep{AndrieuAndMoulines2006}. Figure \ref{fig:spider} shows
the probability density functions for trigonometric parallax produced
with and without the spider mechanism.  Whereas the location of the
histogram's peak changes very little as a result of the spider
mechanism, several chains that were stuck in a local maximum around
290 mas moved to the main peak as a result of employing the spider
mechanism.

Perhaps most importantly, the spider mechanism eliminated the need for
educated guesses as to a starting value for each chain, thus
minimizing the effect of unintentional priors. With the exception of
trigonometric parallax, which was strongly constrained by that of the
$\varepsilon$ Indi A component, the starting values for all other
parameters covered a broader range than those reasonably deduced from
the setup of the observations.

At large step numbers, as long as the algorithm converges, the
  step size described in Equation (A2) stabilizes and becomes
  independent of the current state of the chains, ensuring that our
  MCMC algorithm remains ergodic \citep{AndrieuAndMoulines2006}. We
  demonstrated in Section \ref{subsec:statistics} that our algorithm
  converges, ensuring that the step becomes independent of the chain
  state after the burn in phase.

\subsection{Computational Performance} \label{computation}

The probability density functions shown in Figures \ref{fig:pdf1}
through \ref{fig:pdf3} show the last 1000000 steps of 52 chains of two
million steps each. Because each chain is completely independent from
the other chains the IDL code can be run in parallel by running it in
multiple IDL sessions with identical starting parameters. A procedure
included in the code distribution can be used to easily consolidate
the results from multiple sessions. Running four simultaneous
sessions with 13 chains each and two million steps per chain took
about 10 hours in a MacBook Pro with an Intel i7 dual core processor.

\begin{figure}[h!]
  \includegraphics[scale=1]{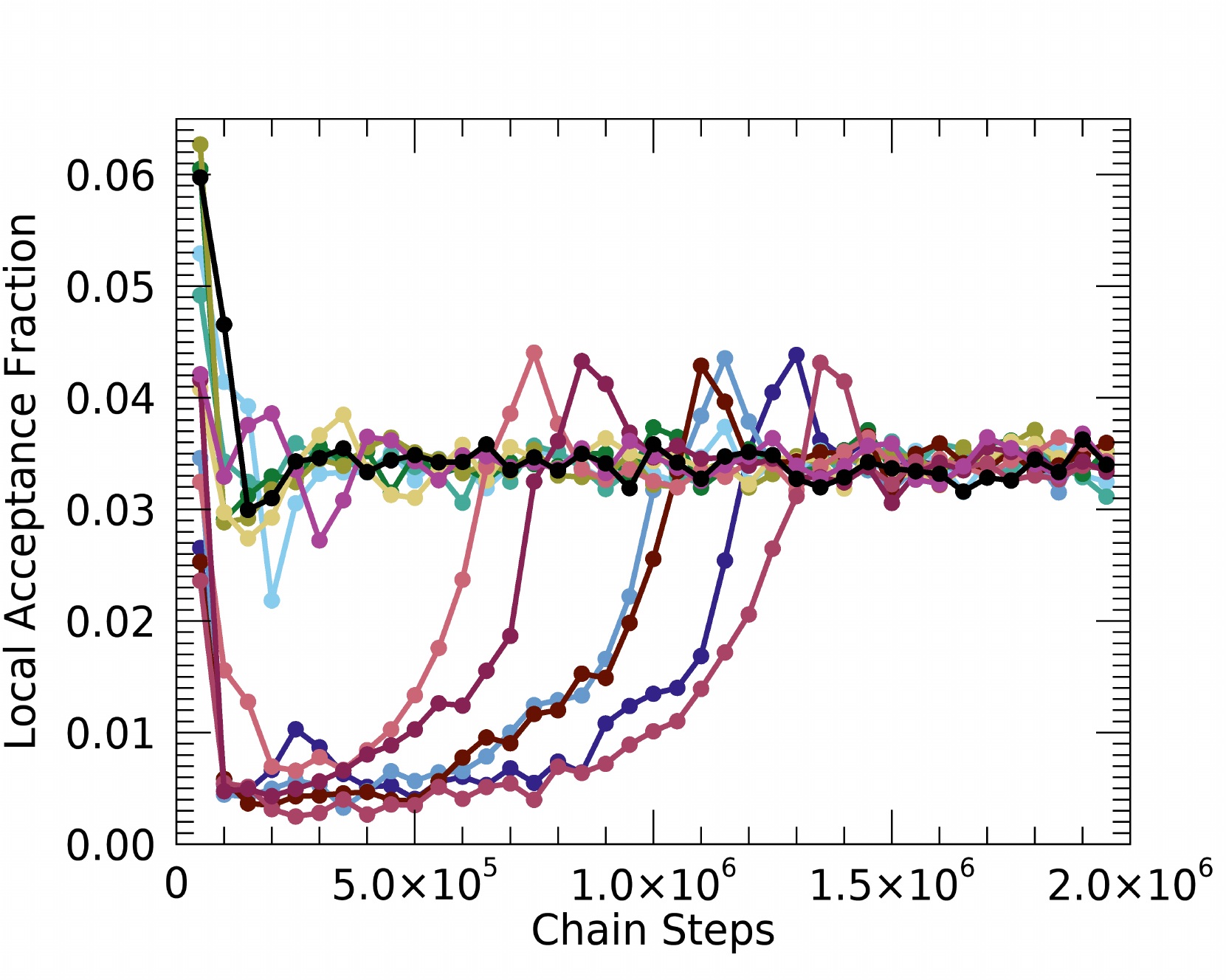}
  \caption{\scriptsize The acceptance fraction for 13 chains, shown as the fraction at every
      50000 interval. The overall low rate may be due to the nature of the parameter space and its
      high dimensionality or due to the choice of step scaling.} \label{fig:acceptance}
\end{figure}

The acceptance fraction as a function of chain evolution is shown in Figure
\ref{fig:acceptance}. The overall acceptance fraction upon convergence is low, at
only about 3.5\%. This low fraction may be due to the choice of step scaling
or simply a result of the complex nature of the multi-parameter space being probed.
The trends in Figure \ref{fig:acceptance} follow the same pattern as those in Figure
\ref{fig:chains}.

\section{Solving for Individual Masses $-$ An Example \label{appendixb}}
\begin{figure}[h!]
  \includegraphics[scale=1]{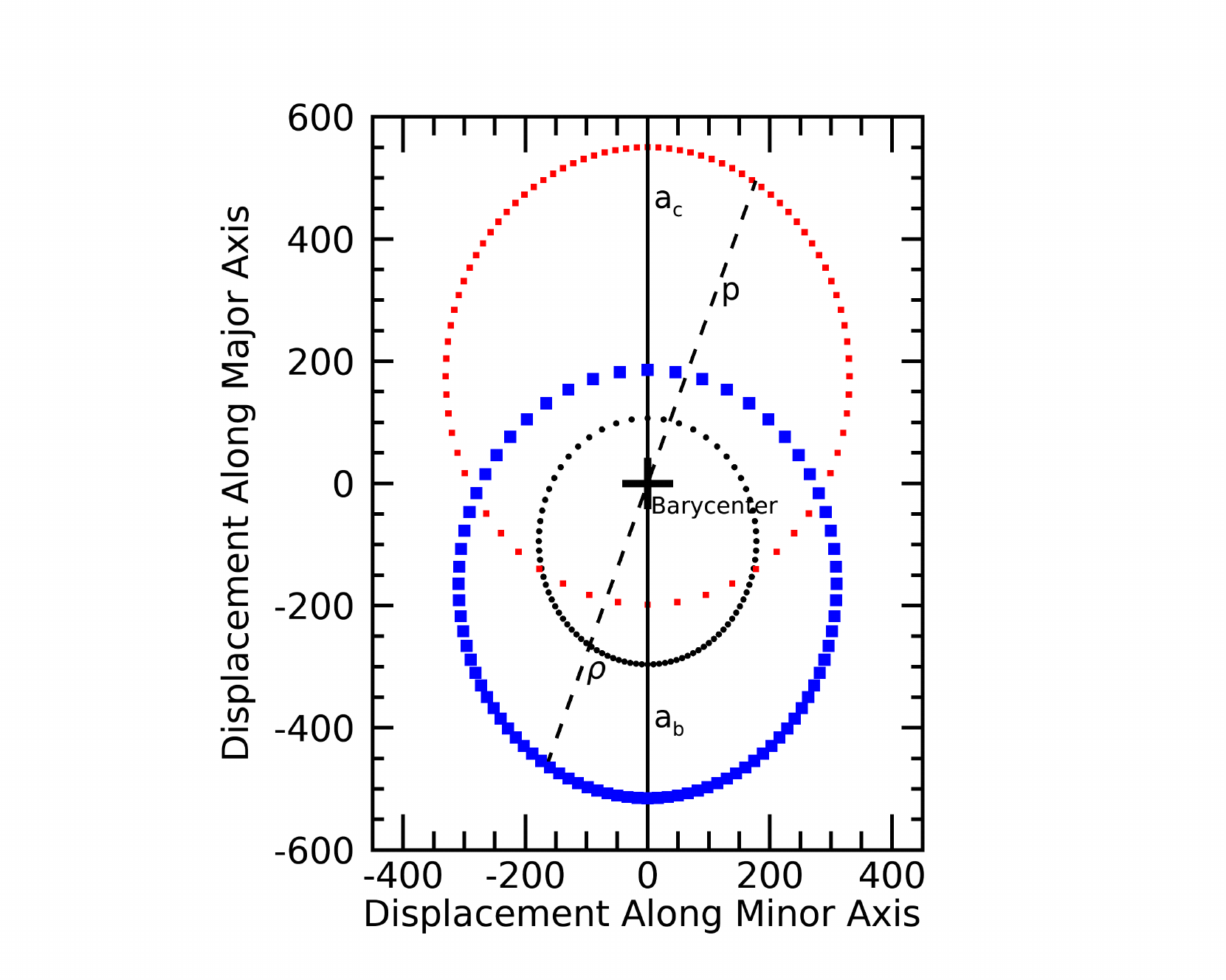}
  \caption{\scriptsize The several orbits involved in the dynamical mass
    problem, shown without projection effects. The large blue squares
    represent the orbit of the more massive components (B) about the
    barycenter. The smaller red squares show the orbit of the C
    component about the barycenter. The black dots trace the observed
    orbit of the photocenter about the barycenter. The dashed line indicates
    the separation between the B and C components measured on August 6, 2005
    (Table \ref{tab:masses}).\label{fig:orbitrelations}}
\end{figure}

We now discuss the problem of solving for dynamical masses given the
photocenter's orbit in detail and carry through a pedagogical example. Figure
\ref{fig:orbitrelations} illustrates the several orbits involved in the
problem as seen from above the orbital plane, with no projection
effects, and with the orbital major axes aligned with the vertical
axis.  These orbits are:
\begin{enumerate}
\item The orbit of the photocenter about the barycenter, traced by black dots. This orbit
  is the one mapped by the astrometric observations and is the starting point for the
  derivation of masses. Here we have a choice of using the photocentric orbit traced by
  the CTIOPI observations or the slightly larger orbit traced by the CAPS observation. We
  choose the orbit traced by the CAPS observations for reasons that will become clear shortly.
\item The orbit of $\varepsilon$ Indi B, the more massive component, about the barycenter,
  traced by large blue squares.
\item The orbit of $\varepsilon$ Indi C, the less massive component, about the barycenter,
  traced by small red squares.
\item The relative orbit of the C component around the B component is not explicitly shown,
  however this orbit can be traced by assuming the position of component B to be static and
  drawing separation vectors towards component C through the barycenter. 
  This relative orbit is the orbit used in solving Kepler's Third Law to obtain the sum of
  the components' masses.
\end{enumerate}

The solution to the problem lies in the fact that all four
orbits have the same eccentricity and their orientation in space may
change only by 180$^{\circ}$, depending on which point is placed at a
focus. It is our task then to scale the size of these orbits based on
the available data photocentric orbit solution, the resolved epochs
of astrometry, and the system's flux ratio so as to obtain the mass sum and the mass ratio for
the BC system.

The dashed lined connecting the orbit of the B component (blue) to the
orbit of the C component (red) represents the separation between the
components measured with adaptive optics on August 6, 2005 (Table
\ref{tab:masses}). We define this separation as $p$
and define $\rho$ as the photocenter's displacement in its
orbit about the barycenter at the same epoch. The quantity $\rho$ can
be calculated from equations 1 and 2 setting the proper motion $\mu$
and parallax $\Pi$ to zero. The fraction $p/\rho$ is then the constant
scaling factor between the observed orbit of the photocenter and the
relative orbit of component C about component B. One clearly resolved
measurement of the separation $p$ is all that is necessary to
establish this relation. Out of the six separation measurements in
Table \ref{tab:masses} we pick the one from this epoch as an example because it
yields the results that are closest to the final weighted averages.
From Table \ref{tab:masses} and calculating the photocentric displacement
from the model yields $p/\rho = 934.8 \, \text{mas} / 260.1 \, \text{mas} = 3.594$.
From this scaling relation we can then establish the semi-major axis
of the relative orbit based on the observed orbit of the photocenter
and its semi-major axis $\alpha$ (Table \ref{tab:results}),
\begin{equation}
  a = \frac{p}{ \rho } \alpha = 3.594 \times 201.61 \, \text{mas} = 724.586 \, \text{mas}, 
\end{equation}
which divided by the trigonometric parallax yields
$a = 0\farcs72458 /  0\farcs27688 = 2.617 \,\text{a.u.}$
We then use Kepler's Third Law of planetary motion expressed in Solar System
units,
\begin{equation}
  (M_{B} + M_{C}) = \frac{a^{3}}{P^{2}},
\end{equation}
to find the system's total mass: $(M_{B} + M_{C}) = 0.1378 \,
M_{\odot}$.

Two points are worth noting here from an observational
perspective. First, we note that starting from an orbit done entirely
on small telescopes and with seeing limited conditions we were able to
obtain the mass sum with a single high resolution observation.  The
ability to do so greatly facilitates the overall observational plan because time
in high resolution facilities is usually more scarce than time in
seeing limited small telescopes. Second, we note that we obtained the
mass sum with no explicit knowledge of the flux ratio between the
components. That dependence is canceled out in the $\alpha/\rho$ factor
in equation B4. We could just as easily have used the smaller CTIOPI
semi-major axis and obtained the same result. This point is
significant because obtaining the correct
flux ratio in a close binary is often challenging and it is sometimes easier
to infer individual masses through indirect ways based on the mass sum, such as
when the primary mass may be known from a mass-luminosity relation.

Now that the total system mass has been established we turn our
attention to obtaining the mass ratio between the B and C components,
and therefore individual masses. Recalling equation 12, We define the
fractional mass of the secondary (C) component as
\begin{equation}
  \mathcal{M} = \frac{M_{C}}{M_{B} + M_{C}}
\end{equation}
and the fractional flux of the secondary component in an equivalent manner as
\begin{equation}
  \mathcal{F} = \frac{F_{C}}{F_{B} + F_{C}}.
\end{equation}
Because the flux ratio between the two components affects the overall
photocentric displacements that trace the photocenter's orbit, it is
crucial that the fluxes in equation B7 be measured in the same band that
was used for the astrometric observations, or that a reliable color
transformation be used. In most cases the high resolution observation
and the astrometric observations are done in different filters, and
color transformations become relevant.  Here we use the
semi-major axis from the CAPS orbit because in Section \ref{subsec:masses}
and Figure \ref{fig:zcomp} we demonstrated that the effective band of
CAPSCam is equivalent to the z band, for which the VLT/NACO flux ratio
is known (Table \ref{tab:epsind}).

We then come to the equation first introduced as Equation 13,
\begin{equation}
  \frac{p}{\rho} = \frac{1}{\mathcal{M} - \mathcal{F}}.
\end{equation}
This equation states that the fractional mass of the secondary
component and the fractional flux of the secondary component play
opposite roles in displacing the photocenter from the position of the
barycenter. To motivate Equation B8 we refer back to Figure
\ref{fig:orbitrelations}. Consider first the hypothetical case where
the secondary component contributes no light. That is usually the
assumption when searching for astrometric perturbations due to
exoplanets. In that case all light comes from the primary component
and the photocentric orbit (black) and the orbit of the primary
component (blue) become the same. The semi-major axis of the photocenter's orbit $\alpha$
is then the
equal to $a_{B}$, $\mathcal{F}$ is equal to zero, and we recover the usual
relation for the location of the components of a binary system about the barycenter:
$M_{1}a_{1} = M_{2}a_{2}$.  At the other extreme consider the case of
an equal mass and equal luminosity binary where $\mathcal{M} =
\mathcal{F} = 1/2$. The symmetry of the configuration then dictates
that the photocenter is placed exactly at the barycenter,  equation
B8 diverges, and no photocentric displacement exists.  Another
interesting case happens when the asymmetry in mass and luminosity between
the two components is small. A small photocentric displacement may
then be detected and the photocenter's orbit then mimics the other
extreme: a very faint brown dwarf or exoplanet orbiting a much more
massive and luminous star.  Detecting the location of the secondary component in
the high resolution observation or establishing an upper limit for
its flux is critical for breaking this
degeneracy and distinguishing between two very different astrophysical
configurations.

From Table \ref{tab:epsind} and transforming magnitudes into fluxes
$\mathcal{F} = 0.206$.  Using the mass sum determined from Kepler's
Third Law (Equation B5) and the ratio $p/\rho$ it is then trivial to
solve Equation B8 for individual masses, yielding $M_{B} = 0.0710
M_{\odot}$ and $M_{C} = 0.0668 M_{\odot}$. These values are very close to the
final values in Table \ref{tab:masses}, obtained using more data and
Monte Carlo error analysis: $M_{B} = 0.0716\pm0.0008 M_{\odot}$ and
$M_{C} = 0.0669\pm0.00064 M_{\odot}$.

\bibliography{/Users/sergiodieterich/sergesreferences2}

\end{document}